# Spark Discharge Generator as a Stable and Reliable Nanoparticle Synthesis Device: Analysis of the Impact of Process and Circuit Variables on the Characteristics of Synthesized Nanoparticles

Miguel Vazquez-Pufleau*[a], Isabel Gomez-Palos[a,b], Luis Arévalo[a], Javier García-Labanda[a] and Juan José Vilatela[a]

*Corresponding author e-mail: miguel.vazquez@imdea.org

a.      IMDEA Materials, Madrid, 28906, Spain.

b.      Department of Materials Science, Universidad Politécnica de Madrid, E.T.S. de Ingenieros de Caminos, 28040 Madrid, Spain.

## Abstract:

Nanotechnology offers the promise of harnessing quantum properties not available in the bulk phase. These desirable properties are highly dependent on size and composition. Generators that control these variables are therefore essential for progress in the field. The spark discharge generator (SDG) is an outstanding aerosol route for nanoparticle synthesis, which stands out due to its fast kinetics, scalability, high purity, accuracy and reproducibility with the added advantage of allowing the synthesis of nanoparticles of any conducting material. These advantages are a consequence of its vast heating and cooling rates, its intrinsic and easily controllable electronic variables at the reach of a click. However, the mechanistic impact of these variables on the actual aerosol generated is still not fully understood. In this work, we constructed an SDG and systematically studied its behavior with particular interest in the effect that resistance, capacitance, inductance, flow rate, gap separation and current have on the electrical behavior of the spark. Our model system produced primarily Fe and Cu nanoparticles with measured concentrations ranging $5*10^5$ - $2*10^7$ part/cm$^3$, and mean agglomerate sizes of 5 – 80 nm. We discuss how the spark influences particle size and number concentration and provide useful correlations that link dependent with independent variables. Remarkably, a finite resistance produces a maximum in the output of the generated aerosols. This suggests a direct link between RLC properties of the circuit and cabling into the frequency of the spark, and nanoparticle number concentration, indicating potential for exploiting such behavior towards maximizing nanoparticle generation. Furthermore, we discuss a link between spark oscillations and energy release with its consequent aerosol generation.

# Introduction:

An important challenge in nanoscience is the reliable and stable production of nanomaterials at specific sizes and concentrations. Synthesis methods for the generation of an aerosol, including flame pyrolysis [1], furnace pyrolysis [2], spray pyrolysis [3], colloids [4], hot filament[5,6], electrospray [7], evaporation/condensation[8], and the spark discharge generator (SDG) [9], and others [10]. In general, these methods have attractive attributes such as control of particle size, concentration, versatility for using different materials, stability and scalability. Additional treatments or purification steps can improve some of these features. For instance, a differential mobility analyzer (DMA) can fix the monodispersity issues from an aerosol generator. The DMA achieves this by size selecting the aerosol of interest based on mobility before further processing. This tactic has been reported for studies requiring monodisperse aerosols whose original source is a furnace reactor [11] or an SDGs [12]. Although such approach is highly accurate for size selection, it gives particles a charge and limits the number concentration. The charge can be later removed by neutralization, reducing the number concentration even more. For this reason, approaches with intrinsically quasi monodisperse production are highly desired.

The spark discharge generator (SDG) is a method that allows a stable production of high purity nanoparticles with controlled size and narrow size distribution from a wide array of materials. Although SDG typically produces a high number concentration of primary particles that tend to agglomerate and become polydisperse, if turbulent dilution is used right after the spark, it can produce quasi monodisperse nanoparticles [9]. This could be achieved by adding large amounts of gas, or by means of a Venturi pump-critical orifice [13].

SDG generates particles in a similar way to laser ablation but simpler; the method does not require a crucible or expensive precursors, has no melting point limitations and can be performed at atmospheric pressure. The instrument is only limited by the capability of achieving a spark between the electrodes of the target material; thus, all conductive elements from the periodic table and mixtures can be directly synthesized via SDG. Semiconductors can also be used if properly doped [14]. Even ceramics or insulating nanoparticles can be synthesized via SDG if a thin layer of the material is deposited on the tip of a conductive electrode. Metal oxides are best produced by oxidizing the particles of their pure metals after their SDG synthesis [14]. Reports show that nanoparticles of more than 20 materials have been successfully synthesized via SDG [15] including carbon [16] [17], metals [18][19][20][21], metal oxides like CuO [21] and $TiO_2$ [22], and semiconductors [14]. The same SDG setup can be used to synthesize all of the mentioned materials, proving robustness and flexibility.

The SDG method is reliable, versatile, simple, clean and cost effective [23]. Its outstanding features are due to the high quenching rate that guarantees complete coalescence of the primary particles and decoupled particle growth from localized vaporization [24]. The SDG is stable for several days of continuous operation [25] [17]. The process can, in principle, be scaled up to yield production rates of ~g/hr [15][18].

Particles generated via SDG can be used for a myriad of applications. Examples include: assembly of 3D structures with controlled morphology via electrostatic lenses [26], synthesis of novel materials with tuned optical, magnetic, and electronic devices [27], web-like aerogels [10], floating catalysts for producing carbon nanotubes (CNTs) [25][28][29], or other 1-D nanostructures [24][30][31][32] and as standard for mass spectrometry and aerosol sizers [33]. These nanoparticles also find application in the fields of radioactive labeling, health, environment, material science, electronics, catalysis, etc. [12][34][35].

In this work, we have successfully tested this device for the synthesis of Ni, Zn, Fe, Cu, Mg and Au with size distributions ranging 1 - 150 nm and concentrations above $10^8$ part/cm$^3$. These nanoparticles are for example suitable as catalyst nanoparticles for growing 1D nanomaterials by the VLS mechanism [31,32][36].

**From High voltage to nanoparticles**

Electrical discharge regimes are classified in increasing order of energy per pulse involved as corona discharge (< 20 µJ), glow discharge (20 µJ - 200 µJ) and arc discharge (>200 µJ) [37] [38]. The third one is generally the most effective for nanoparticle production and is the focus of this work. The spark discharge is generated when the voltage between anode and cathode is sufficient to produce a plasma through which electrons can flow [39]. Once the plasma channel is formed, the spark follows. In the few microseconds it lasts, it generates soft X-rays with energies between 2 and 10 keV [12], and reaches temperatures between 20 000 and 30 000 K [40], immediately evaporating the surface where the spark hits. This causes a plume of the electrode material to evaporate from the electrodes. Hydrodynamic effects dominate during the first few microseconds after the spark [40], and cause the evaporated electrode material to condense forming a high concentration of primary particles around 1 to 10 nm [27]. Then, particles grow and agglomerate further based on aerosol dynamics. Sparks quench due to radiation and adiabatic cooling at rates ranging $10^7$ - $10^{10}$ K/s [41][42]. Below the evaporation temperature, thermal conduction dominates. Aerosol models exist to estimate how particle size distribution and number concentration change over time. These models generally assume that there are collisions between atoms and particles originating from the metal vapor with 100 % sticking efficiency in an ideal Smoluchowski type of coagulation process [15]. This approach provides a good first estimate of the dynamic aerosol evolution, although reports show these assumptions are not always valid under SDG conditions [2]. The nanoparticle mass throughput is ~5 g/kWh.

**Mathematical relationships for SDG**

This section summarizes useful relationships to estimate particle formation as a function of spark conditions. In general, the spark characteristics are very sensitive to the gas composition, the shape of the electrodes, and the presence of electronegative molecules. Therefore, it is more accurate to measure them than to use only theoretical calculations [18]. In any case, a helpful starting point to estimate breakdown voltage is the Pashen´s law:

$$V_b = \frac{BPd}{ln\frac{APd}{\ln(1+1/\gamma)}} \qquad \text{Eq 1}$$

, where **A** and **B** are constants, **γ** is the Townsend secondary ionization coefficient, **P** is pressure and **d** is the distance between electrodes. For air at standard temperature and pressure (STP), **A**= 12, **B**= 365 and **γ**= 0.02. Breakdown voltages predicted with Paschen's law have been reported for different gases as a function of pressure and gap distance [12][18]. The higher the pressure, the higher the breakdown voltage

($V_b$), and the same happens for gap distance, so both variables are interconnected. For example, Ar has a higher $V_b$ than He, providing higher energy per spark in Ar than in He for the same $P$ and $d$. Since the slopes of $V_b$ are steeper for air and $N_2$ than for He and Ar, they are more strongly affected by the inter-electrode gaps and gas pressures [18]. The higher the energy put into a spark, the more material gets vaporized. The energy released per spark is given by:

$$E = \frac{1}{2}CV_d^2 \qquad \text{Eq 2}$$

, where $C$ is capacitance, and $V_d$ is the discharge voltage. A problem occurs with the discharge of the classical circuit for spark discharge. $V_d$ is not constant and is higher than the breakdown voltage $V_b$:

$$V_d = V_b + V_0 \qquad \text{Eq 3}$$

, where $V_0$ is an overvoltage that is irregular in nature. At a fixed electrode gap distance and a constant capacitance, the frequency is given by the constant current $I_C$ charging the capacitor and the voltage at which the discharge occurs:

$$I_C = C\frac{dV_d}{dt} \qquad \text{Eq 4}$$

The mass loss rate from the electrode depends on the repetition frequency ($f$):

$$f = \frac{I}{CV_d} \qquad \text{Eq 5}$$

, where $f$ is the frequency of discharge in Hz, and $I$ is the constant current applied to charge the capacitor. Since the values of $V_0$ are somehow unpredictable, the measurement of frequency is more recommended than its calculation [18]. The actual time of gas breakdown (conducting channel formation) is sensitive to the electric field, and thus to $V_d$. Ionization is also dependent on gas temperature and pressure, on the field configuration, the electrode surface and gas composition including impurities. Garwin et al. 1988 reported an equation to describe the concentration of particles $n$, produced in an SDG when no coagulation occurs [9]:

$$n = N_0\frac{f}{\dot{V}} \qquad \text{Eq 6}$$

, where $N_0$ is a constant number of primary particles formed per spark discharge, and $\dot{V}$ is the volumetric flow rate. For large aerosol concentration, when coagulation is strong, losses to the walls and turbulent dilution can be neglected because they are linearly proportional to the concentration, whereas coagulation is proportional to concentration square [24]. If particles are spherical and have a chance to sinter, one gets the following expression for the dynamic evolution of particle number concentration as a function of coagulation [24]:

$$\frac{dN}{dt} = -\frac{1}{2}\beta N^2 \qquad \text{Eq 7}$$

, where $\beta$ is the coagulation kernel and depends on temperature and particle size distribution, gas flow conditions and inter particle forces [24]. Solving for $N$ at final time ($N_{(tf)}$) and considering that residence time of the particles is $t_f = V_{eff}/Q$, one gets:

$$N_{(tf)} = \frac{2}{\beta V_{eff}}\dot{V} \qquad \text{Eq 8}$$

, where $\dot{V}$ is the gas volumetric flow rate, and $V_{eff}$ the effective volume occupied by the coagulating aerosol [24]. Thus, particle size (**dp**) becomes:

$$d_p = \left(\frac{3\beta V_{eff}\dot{m}}{\rho\pi Q^2}\right)^{1/3} \qquad \text{Eq 9}$$

, where $\dot{m}$ is the mass production rate. **β** can be approximated to be constant without much error. Values of **β** = 3.29*10$^{-16}$ m³/s and $V_{eff}$ of 66 cm³ have been used for 5 nm particles and considering the duct volume and its aerosol plume [24]. The mass produced at a single spark is:

$$\Delta m = C(E - E_0) \qquad \text{Eq 10}$$

, where **C** is a constant that depends on material, **E** the spark energy and **E₀** the minimum spark energy required to form particles. It follows that:

$$d_p = \left(\frac{3\beta V_{eff}C_{tt}(E-E_0)f}{\rho\pi Q^2}\right)^{1/3} \qquad \text{Eq 11}$$

, where **ρ** is the material density, and $C_{tt}$ is a material dependent constant, given by [24]:

$$C_{tt} = \frac{\alpha}{c_{ps}(T_m-T)+c_{pl}(T_b-T_m)+H_m+H_e} \qquad \text{Eq 12}$$

, where $T_m$ is the melting point of the electrode, $c_{ps}$ and $c_{pl}$ are the heat capacities of the solid and liquid material respectively, $H_e$, and $H_m$ are the enthalpies of vaporization and melting respectively, and **E₀** is given by:

$$E_0 = \frac{2\pi r\tau[r\sigma(T_b^4-T^4)+k_e(T_b-T)+k_a(T_b-T)]}{\alpha} \qquad \text{Eq 13}$$

, where **r** is the radius of the ablated hot-spot by the spark, **τ** is the spark duration, **σ** is the Stefan Boltzmann constant (5.67*10$^{-8}$ Wm$^{-2}$K$^{-4}$), **T** is the gas temperature, $T_b$ is the boiling temperature, $k_e$, and $k_a$, are the thermal conductivities of the electrode and carrier gas respectively, and **α** is the fraction of the energy used to evaporate the electrode. An expression to relate the physical properties of the electrode material to its erosion and consequent mass evaporated was proposed by Lleweyn Jones et al. [43]. It seems to be a good starting point for estimations but some discrepancies between experiment and model have been reported for some systems [18].

$$m = \frac{\frac{1}{2}c_eV^2-bT_b^4-gk(T_b-T)}{c_{ps}(T_m-T)+\Delta H_m+c_{pl}(T_b-T_m)+\Delta H_v} \qquad \text{Eq 14}$$

, where $c_e$ is the gap capacitance **V** *is* the breakdown voltage, $T_b$ is the boiling point and **b = Aσt (J/K$^{-4}$)**, **σ** depending on the blackness of the object. For black bodies **σ** is Stefan Boltzmann constant (5.67 *10$^{-8}$ J/(s m² K⁴). The time for energy transfer is **t**, **k** is thermal conductivity (W/mK) and **g = 2(πA)$^{0.5}$t** (ms). If **r²/Kt)$^{0.5}$** << 1 in which **K** is the thermal diffusivity (m²/s) and **r** is radius. $T_m$ is the melting point of the electrode, $c_{ps}$ is the average heat capacity and **ΔH$_m$** is the enthalpy of melting. $c_{pl}$ is the heat capacity of the liquid, and **ΔH$_v$** the enthalpy of evaporation. The mass evaporated for each spark is between 0.18*10$^{-6}$ and 1.5*10$^{-6}$ mg [25].

**Implications of this work**

Desirable properties in nanoparticle applications are a strong function of particle size and concentration [44]. Therefore, it is critical to characterize the variables from which it depends to enable its deterministic synthesis. This study attempts to contribute in these regards.

Several studies [12] [15] [45] report improvements in the SDG since its first report by Schwyn et al. [9]. Unfortunately, it is not yet fully understood how the different components of the electric system affect the spark [12], and consequently, the nanoparticle properties such as size and concentration. Regarding the RLC circuit in an SDG, the capacitance effect has received some attention [9] [17] [18][44], but the effect of inductance and resistance is still not available in the literature [12]. A systematic investigation is necessary in order to fully understand its principles and how the different electrical parameters have an influence on the spark, charging and aerosol formation [12]. In the current study we asses these influence by comparing our measured spark features at different conditions with the trends predicted by Eq. 1, 2 and 4 and found unexpected correlations for the effect of resistance.

Additionally, some reports show that a single spark when measured with µs resolution, displays electric oscillation [46], [15]. Such electric oscillations are linked with the energy released per spark and consequently with the nanoparticle production. Therefore, looking into the electric behavior with such time resolution, indirectly provides the potential for monitoring the aerosol formation. For this reason, in this work, we studied systematically the sub-µs resolved voltage variation in sparks, and performed current analysis aided by Ohms law. We correlate the dynamic behavior as a function of independent circuit variables such as current, flow rate, gap distance, resistance, capacitance and inductance. We also correlate the variables with their effect on size distribution and total concentration. For clarity and conspicuity, the results for each variable are discussed in a dedicated subsection in this paper. Aerosol sizes reported refer to agglomerate sizes as measured by a scanning mobility particle sizer (SMPS).

## 1. Experimental:

Our SDG was constructed based on a combination of standard equipment and custom-made components to achieve implementation flexibility. Fig. 1a shows a schematic of the setup with sampling and ancillary elements. Fig. 1b displays the circuit and voltage characterization instruments. Fig. 1c displays a spark between the electrodes. The energy for the spark was delivered by a PNC 20000-10 UMP direct current (DC) high voltage power supply (Heinzinger, Germany) with a range of 0 – 10 mA and 0-20 kV. In this study we used positive polarity. After the power supply, the high voltage passed through an electronics box for its processing through a NTE518 diode (Farnell, USA), resistances, capacitors and inductors (RCL) system. Afterwards, the current was transported towards the spark chamber by means of an 8-meter long shielded Heinzinger® HVC30 cable (Heinzinger, Germany). This long cable was necessary for practicality reasons because the SDG was built to feed of nanoparticles our 1-D nanomaterial reactors, and the high power supply components are heavy and the nanoparticles generated by the SDG had to be produced as close as possible to the reactors inlet several meters above ground level. Therefore, a portable setup was the best option for these experiments. The electrodes used were 3 mm in diameter and their purities include 99.9 % for Cu, 98% for Fe (RS Components, United Kingdom), and 99.95 % for Au (Goodfellow, United Kingdom). The electrode distance is manually adjustable by means of an XYZ micrometric sliding table. The continuity option of a multimeter was used for zeroing the electrode gap separation by reducing the distance with the micrometer knob until circuit continuity was measurable.

Unless otherwise specified, N$_2$ 99.999% purity was used (Airliquide, France). To feed the SDG, a FMA5520A mass flow controller (Omega, USA) was used, with a working range between 200 sccm and 10 slpm of N$_2$, connected to a homemade potentiometer with accuracy up to +/- 10 sccm. The SDG chamber is about 1 liter.

To measure high voltage with an oscilloscope, a CT2982B high voltage probe (Cal Test Electronics, USA) with 1000-X reduction capabilities and acquisition speed up to 40 MHz and 10 kV with 3% accuracy was employed. This probe delivered safe voltage to a Picoscope 2406B oscilloscope (Pico Technology, United Kingdom) with up to 1 ns time resolution and accuracies of 3% in DC and 4% in AC. Picoscope 6 software was used to acquire the time-resolved data. A self-developed Python program was used to average 32 measurements of single sparks for each reported condition.

Transmission electron microscope (TEM) images were acquired by means of a field emission Thalos 200 TEM (FEI Company, USA), scanning electron microscope (SEM) images were obtained via a Helios nanolab 600i focused ion beam SEM (FIB–SEM) (FEI Company, USA), and energy dispersive X ray spectroscopy (EDX) analysis was performed via Aztec (Oxford Instruments, United Kingdom).

For acquiring aerosol size distributions in real time, an SMPS+E (GRIMM Aerosol Technik Ainring GmbH & Co. KG., Germany) was employed. The SMPS counted with a GRIMM 5706-230/50 DMA and flow controller unit, a 5705 Faraday cup electrometer (FCE) provided the counts of the negatively charged particles in each bin of the aerosol size distribution and were acquired via a 5477 GRIMM nanoSoftware. The aerosol flow was charged with a 5524-X soft X-Ray neutralizer from the same manufacturer. An M-Vienna-Type DMA with an effective length of 88 mm was used for size selection in terms of mobility diameter. The sheath flow rate used was 10 lpm and the aerosol flow rate 1 lpm. With these settings, the measurable size distribution range was 2.76 - 153.9 nm. The integrated correction parameters for multiple charge, DMA efficiency correction, FCE offset correction and Inlet loss corrections available in the GRIMM nanosoftware were utilized for all SMPS characterizations in this work. Such corrections of size distribution follow recommendations from ISO 15900. Several of the number concentrations scanned have a section which lies close to the limits of experimental detection by an SMPS (<2.76 nm) and are also limited by intrinsic temporal aerosol dynamics, as shown by Fig S2 of [24]. However, this limit is far away from the mode of all size distributions measured and its effect on the mean and total aerosol concentration is believed to be small.

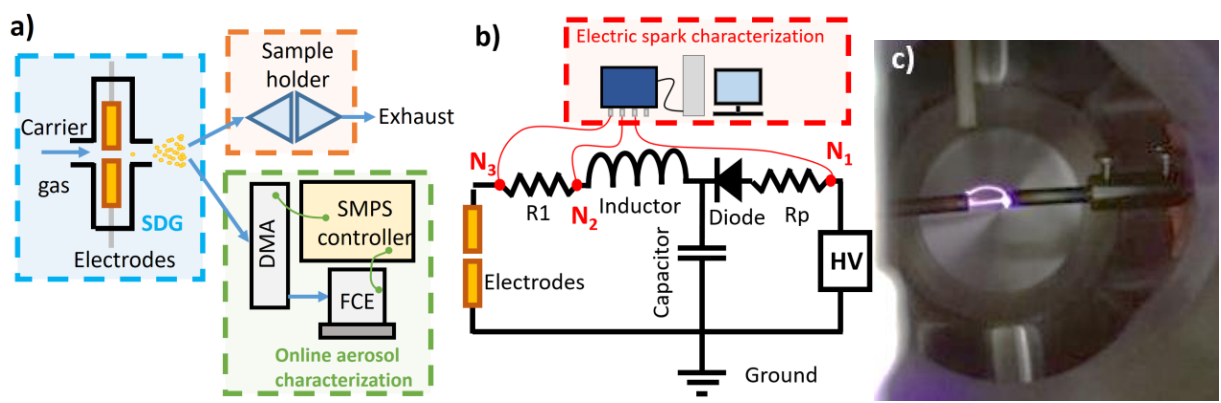

Figure 1. a) Flow schematic of the apparatus including the SMPS characterization instrument and sample holder which can host a filter or a TEM grid for offline characterization. b) Electronics diagram of the spark discharge generator. *R1* is the resistance added to the circuit for measuring voltage drop and allowing the determination of current via Ohm´s law. *Rp* is the resistance added to the circuit to protect the power supply against inverse voltage. The electronic behavior of the SDG was observed by connecting a voltage-measuring device in one of the 3 nodes $N_1$, $N_2$ and $N_3$ and recorded via an electronic device. c) Picture of an actual spark between the electrodes.

Table 1 Experimental plan and Figures showing their results

| Independent variable | Section | Spark analysis | Current analysis | SMPS | SEM/TEM | Gov. Eqs. |
|---|---|---|---|---|---|---|
| Resistance before diode | 2.1 | Fig. 2, S1 | | | | |
| Current | 2.2 | Fig. 3 a,b, S5 | | Fig. 3c,d,e | | 4,5 |
| Gap separation | 2.3 | Fig. 4a,b, S6 | | Fig. 4c,d,e | Fig. S3 | |
| Gas flow rate | 2.4 | Fig. 5 a | | Fig. 5b,c,d | | 8,9,11, |
| Resistance before spark | 2.5 | Fig. 6a, h, S8 | Fig. 6d-h. S9, S10 | Fig. 6b,c | | |
| Capacitance | 2.6 | Fig. 7a,b | Fig. 7c - g, S11 | Fig. 7h, i | Fig. S2 | 2,4,5 |
| Inductance | 2.7 | Fig. 8a, S12 | Fig 8d-h, S13-15 | Fig. 8b,c | | |
| Electrode material | | | | | Fig. S2 | 12,13,14 |

## 2. Results & discussion

## 2.1. Design considerations

**Power supply reverse voltage protection.** Spark discharge generally has the drawback of back current, which can damage the electronics and the power supply that produces it. In this study, to protect the high voltage generator from back current, a diode was included between the generator and the capacitor. However, the diode was too slow and finding a high voltage and ultrafast diode proved challenging (desired response time <100 ns), so we decided to use a resistor between the power supply and the circuit. Such approach has been chosen by other researchers for SDG circuitry [25][19][15][22][40][30][47][48]. However, it is important to utilize an appropriate resistor, as an excessively large resistance would cause unnecessary losses by power dissipation ($P_{loss}$), and a too small one might not protect the power source sufficiently. $P_{loss} = V*I$, by Ohm´s Law $V = I*R$, so $P_{loss} = I^2R$ (Fig. 2a). Therefore, the importance of the analysis. Disagreement seems to exist in the literature on the best way to proceed with the resistance for protecting an SDG high voltage power supply (HVPS). Some authors report using a fairly small or no resistance: 0 and 5 Ω [15], 150 Ω [19], 220 Ω [30]; whereas others use substantial resistance values of 500 KΩ [40], [47], 1 MΩ [25], 1.5 MΩ [48], and 5 MΩ [22].

For the conditions tested of $G$ = 1.5 mm, C = 12.47 nF, $\dot{V}$ = 1.3 lpm $I$ = 1mA, $L$ = 100 µH (Fig 1b), the lowest resistance that prevented negative voltage from reaching the power source was between 2 and 3.3 kΩ (Fig 2 b). In a further test, a 3 kΩ resistor was scanned and showed the same level of protection against negative voltage. However, the voltage at $N_1$ (before the diode) still depends on the following variables as further mentioned (Fig. S1):

- **Current** does not change the voltage behavior of the spark when measured at the node before the diode ($N_1$) in a similar way as happens before the spark ($N_3$).
- **Flow rate**. The higher the flow rate, the higher the breakdown voltage, but the lower part of the ripples of high flow is farther away from reaching zero than the lower flow cases (1 lpm touches the 0 V line).

- **Gap separation** shows the same type of behavior than flow rate. The higher the gap, the higher $V_d$ in the spark, and therefore, the farther away the ripples from the zero gap are.
- **Capacitance** does not affect $V_d$, but the ripples at higher capacitances go closer to zero than at lower capacitances. In addition, the amplitude of the oscillations increases at higher capacitances.

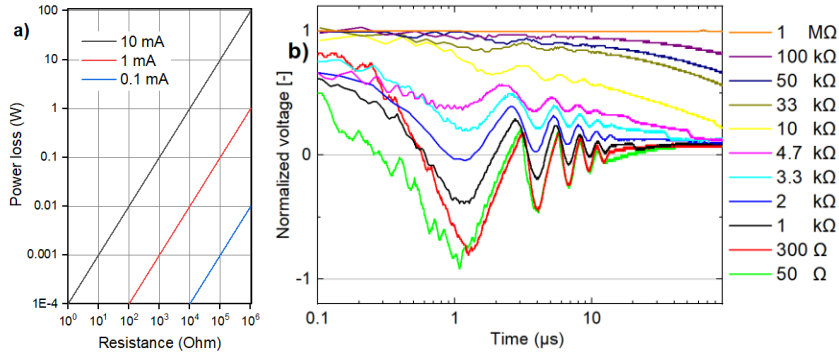

**Figure 2. a)** Power loss at different protection resistances ($R_p$). For scaling up its dependence to $I^2$ makes $R_p$ selection critical. **b)** Effect of resistance in normalized voltage before diode ($N_1$ in Fig. 1). Conditions were: $G$=1.5 mm, $C$=12.47 nF, $\dot{V}$=1.3 lpm, $I$=1 mA, $L$=100 µH, $R_{al}$=none, $E_m$=Fe98%. The effect of capacitance, gap distance, current and flow rate can be seen in Fig. S1.

**Primary particle vs agglomerate size.** The size distribution of aerosols generated by spark generators is near-lognormal, with standard deviations around the self-preserving value of 1.46 and below for distributions containing a mean size smaller than 3 nm [15]. The primary particles are usually quite spherical. The primary particle size is taken as the largest coalesced unit [2][49]. Extra care for characterizing these small particles is required as they might coalesce to a larger size under the electron beam of TEM, or even at room temperature. In SDG, the energy per spark determines the primary particle size. These primary particles tend to stick together and agglomerate above certain concentrations due to Van der Waals forces. A critical flow rate exists, above which particles do not agglomerate and below which they do. The best way to control primary particle size distribution, which can be more important than the agglomerate size for some material properties [2][50], is either through the energy per spark or through turbulent dilution near the spark. A comprehensive study on primary particle size analysis is beyond the scope of this study, but some preliminary TEM images are available in Fig. S2 and S3. To later turn agglomerates into fully compacted spherical particles, a furnace can be used after the spark generator [23]. If purity is a concern, an extra careful operation is necessary as most furnaces release trace amounts of materials that adsorb in the aerosols [15].

**Energy per spark.** Agglomerate size and concentration are directly given by the energy of each spark. However, the best approach to achieve non-agglomerated primary particles is to increase the flow rate beyond a threshold. For instance, by blowing a flow directly into the spark, particles of 1 nm can be generated. Schwyn et al. achieved this by using around 23 mJ per spark [15]. Horvath et al. [17] observed mass production rate by DMA and gravimetrically. They report that it starts to plateau above ~12 mJ per spark. Their conditions were $C$=6 µF, $V$=150 V, $N_2$ or Ar, and $G$=1 mm (although their system was gap independent, as discussed previously). On the other hand, Pfeiffer et al. [15] observe a fairly linear behavior in terms of electrode mass loss up to 25 mJ/spark with conditions: $G$=1mm, $C$=20 nF, $\dot{V}$=0.8 lpm

Ar. The difference might be attributed to one of the cases approaching aerosol concentration saturation but not the other [51].

As Eq. 2 indicates, there is a direct relationship between energy per spark, breakdown voltage and capacitance. The primary particle size is directly linked to energy per spark (Eqs. 8 and 10). However, the voltage is not directly controlled in our SDG, it mainly depends on the carrier gas used, gap separation and flow rate. A discussion on the effect of these independent variables in the spark is presented in the next sections. The energy per spark in our equipment was estimated to be 40 mJ using Eq. 2 and 100 mJ/spark by doing circuit analysis and integration.

**Nanoparticle production rate.** The yield of our SDG was estimated based on the mobility diameter reported by SMPS. For the calculation, we assumed spherical particles and obtained 10 mg/hr of nanoparticles, using the following conditions: $G$=5.5mm, $C$=10nF, $\dot{V}$=3 lpm, $I$=10 mA, $L$=1 µH, $E_m$=Fe. For different conditions, namely, ($G$=2mm, $C$=32nF, $\dot{V}$=20 lpm, $I$=10mA, $L$=1 µH, $E_m$=Fe, Zn, Mg, Cu), we achieved production rates of: 1.2 mg/hr Fe, 8 mg/hr Zn, 4.5 mg/hr Mg and 1.5 mg/hr Cu, as measured gravimetrically on the nanoparticles collected in a filter. This correlates well to reported yields of 43 mg/hr using 17 kV, Au + Ag in one electrode and W in the other, 45% power, $G$=3 mm, $f$=480 Hz, $\dot{V}$=1.2 lpm Ar [52], and also with the 11 mg/hr of Tabrizi et al. [18] using $f$=150 Hz, $C$=20 nF, G=1 mm and $\dot{V}$=0.8 lpm Ar [18].

Based on calculations, about 0.15% of the energy gets used to produce nanoparticles. For $G$=1mm and an electrode diameter of 6mm, 1 g of Au nanoparticles could be produced from $10^6 – 10^7$ J. An SDG capable of producing 25 mJ sparks at 25 kHz would produce 1 g of gold in about 27 min for each pair of electrodes [15]. Horvath et al. report that mass concentration increases exponentially up to 10 mJ and then just marginally, with 7 mg/hr as their maximum [17].

**Losses through distance.** In general, the shortest the distance between an aerosol generator and its characterization instrument, the better. In our case, our SDG was built to feed a reactor for 1-D nanomaterials synthesis and it was more relevant to know the nanoparticles being delivered to the reactor rather than right after the spark. Due to the bulkiness of the SDG and reactor itself, 2 m was considered the shortest distance between the two. For this reason, the aerosol characterizations reported in this work were executed after passing through a 2 m SS ¼" pipe from the spark. Calculations show sedimentation losses across this pipe are negligible for our experimental conditions. Furthermore, a study reports that the concentration reduction for a comparable distance is around 15% [52]. To minimize losses, it is recommended to locate the generator as close as possible of its intended operational position.

**Charge of particles.** The spark generator produces multi-charged particles with both polarities. SDG particles could be introduced directly into a DMA, but the measured size distribution would be wrong since the DMA relies on a bipolar neutralizer to provide particles with a charge of zero or unity at Boltzmann equilibrium [53]. In our case, the measured Fe nanoparticles concentration was one order of magnitude higher when using the neutralizer, but the size distribution did not change, in agreement with previous studies [24]. For Ni nanoparticles, the shift in size distribution with and without a neutralizer was more pronounced. For these reasons, a neutralizer was used for all reported DMA characterizations in this study.

In SDG, material from one electrode can form a jet and travel through the gap, eroding the opposite electrode. The cathode tends to erode more than the anode due to the higher mass of positively charged ions compared to the negatively charged ions (electrons) hitting the anode [18]. To balance the losses in both electrodes, the polarity of the system can be changed. However, the system might cause the spark to change the electrode polarity several times within a single circuit discharge, as we explain below.

The high proportion of charged particles going out of the spark generator can be a great advantage as other techniques have trouble charging small clusters. Generally, there are about twice as many negatively than positively charged particles [15]. Negative particles reach up to 20% of the total amount of particles. Positive particles are less abundant perhaps because the negative electrode is grounded together with the housing, causing polarity dependent wall losses where negative particles have a higher survival likelihood [18]. In carbon nanoparticles, 90% of 20 nm particles are charged but only 50% of 50 nm particles are. Particles smaller than 10 nm are reported to be close to 100 % neutral [54]. But this does not mean charged particles smaller than 10 nm are not existent. Maisser et al. have been able to measure even sub-nm charged clusters from an SDG using a high resolution DMA [55].

**Ejection of micrometer spherical particles.** An unusual mechanism for the formation of large particles occurs in some operating conditions of the SDG. If the recoil force (explosive evaporation or acoustic ejection) of the spark is larger than the surface tension, a droplet of molten electrode can be ejected. This phenomenon produces comparatively big spherical particles ranging from a few nm to several µm in size [14] [18][56][57][58][59]. We also observed the formation of these large particles (See Fig. S2b encircled in red). The generation of these large particles seems to be favored by conditions associated in this study with high energy, in agreement with [14]. This phenomenon might still happen for less energetic sparking conditions if the spark consecutively hits the same spot [15]. The formation of these µm particles might be desirable or undesirable based on the application and should be taken into consideration for optimizing the SDG, but it is beyond the scope of this work.

**Cable between RLC circuit and spark**. An 8 m cable was used in this study to connect the RLC circuit to the electrodes (the actual place of sparking). The cable has an intrinsic capacitance of 127 pF/m and an inductance of 0.39 µH/m, with an impedance of 50 Ω, so the effective cable capacitance is around 1 nF and the inductance 3 µH. All the calculations based on Eq. 5 systematically deviated between prediction and experiment. By assuming an additional system capacitance of 7 nF the deviation could be reasonably fixed for all uses of Eq. 5. Since such capacitance can be attributable to the cable, circuit and HV power source, we use it through this paper.

**Figures of merit description**. From this section onwards, most Figs. were prepared from systematically acquired data scanned across one variable at a time. A description of these figures follows. The upper and lower part of *V* just before and after the spark, its standard deviations (Fig. S4), as well as *f* for all variables is estimated using Eq. 5 and compared to the measurements (a). Time-resolved voltage dependences for the variable in question were obtained by integrating 32 measurements and are displayed in 3D-plots and some of them in 2D plots in the SI.

## 2.2. Current

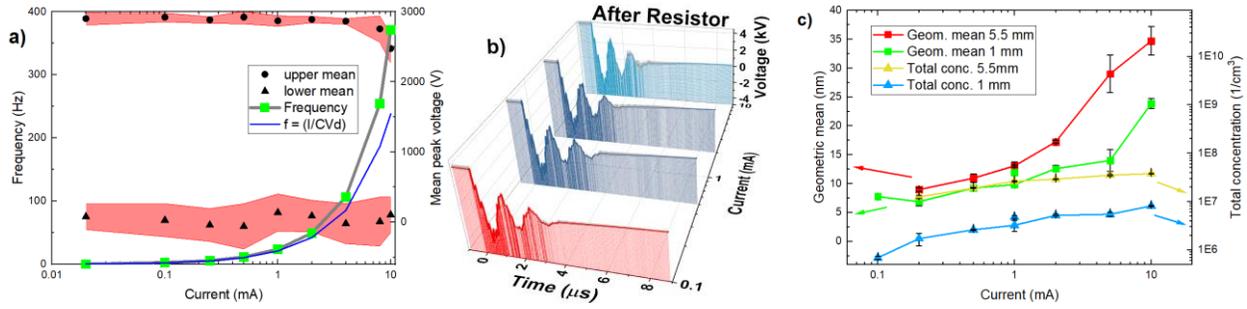

**Figure 3. a)** Effect of current in the system for frequency and voltage observed in the node 3 ($N_3$) just before the spark. Bounds of the standard deviation of the peaks are represented by the red band around the mean values of upper and lower voltage measurements of numerous sparks at the given conditions. Fig. S5 shows the effect of current on frequency for different capacitances. In general, the observed voltage does not change significantly for different currents and only the frequency is affected. The conditions used were $G$=0.5 mm, $C$=10 nF, $\dot{V}$=1 lpm $N_2$, $L$=100 µH, $Em$=Fe. **b)** The shape of the spark also does not seem to change too much as a function of current. Used conditions: $G$=1 mm, $C$=2 nF, $\dot{V}$=2 lpm $H_2$, $L$=1 µH, $R_{ai}$ = 1 Ω, $Em$=Au. Also taken at $N_3$. **c)** Geometric mean and total concentration obtained via SMPS respectively. The conditions used were $G$=5.5 mm and 1 mm, $C$=2 nF, $\dot{V}$=3 lpm $N_2$, $L$=1 µH, $Em$=Fe. Size distributions can be found in Fig. S5a.

Current (*I*) is the best variable to modify frequency without changing discharge voltage (*$V_d$*) (Fig. 3a). There is a slight decrease of *$V_d$* at higher currents, perhaps because above 10 mA the spark becomes continuous. Eq. 5 *(f= I/CV_d)* describes qualitatively the relationship between *I* and frequency (*f*), (left axis in Fig. 3a and Fig. S5b). If only the nominal capacitance (*C*) of the capacitor and the one from the 8 m HV cable were considered, *f* would deviate systematically suggesting *C* underestimation. Therefore, we used the capacitance correction mentioned in the previous section. The shape of the spark seems to be unaffected by *I* (Fig. 3b). The increase in *f* is in agreement with previous studies [16][18].

There is a limitation on the maximum *f* that can be achieved by a standard SDG. This is reportedly around *f*=500 Hz, but can be increased to 1 kHz with capacitor charging supplies with large output resistors [15]. We did reach *f*= 1200 Hz with our standard SDG but at very short electrode gaps (*G*) and low *$V_d$* (Fig. 4a). A typical spark lifetime is about 10 µs then there is about 1 ms wait time for the circuit to recharge. Therefore, a limit of sparking at 100 kHz might exist. However, a practical problem using a standard SDG above 200 Hz is that the setup runs into a transition regime due to arc discharge where *$V_d$* diminishes because charges from the previous sparks allow premature sparking before the capacitor has fully charged to its desired level. Since energy in the capacitor scales with *$V_c^2$*, this affects particle formation significantly. Such problem can be solved if the charge and discharge are decoupled by using switches, so the discharge is not limited by the breakdown voltage of the gas but rather by the switch trigger. To avoid randomness (*$V_0$* of Eq. 3), a continuous but low power glow current could be kept [15].

A systematic exploration of the current effect in the SDG led to the following findings: a) higher *f* translates into higher mass throughput (Fig 3c), in agreement with [10]. b) both particle concentration and mean particle size increase at higher *I* (Fig. 3c), in agreement with [44]. c) this effect (see the slope in Fig. 3c) is more pronounced at higher gap separations (5.5 mm) than at lower ones (1 mm).

Our observation is in contrast with some reports: Krasnikov et al. [60] indicated that a current increase affects yield but does not change the mean particle size [60]. Our results do show that agglomerate sizes grow as a function of *I*. Perhaps this discrepancy is due to the narrow range of currents tested by Krasnikov (0.4 - 0.8 mA), where the growing agglomerate effect might not be strong enough for easy visualization (not even for us in such range). Another study by Tabrizi et al. [18] reported that particle concentration first rises with *f* but decreases at higher *fs*. However, we observe a monotonic increase in particle size for all *I*s tested. This discrepancy might be due to Tabrizi et al. modifying other variables in addition to current to change *f*, such as $V_d$, *C*, or having reached saturation concentration, whereas we do not.

## 2.3. Gap separation

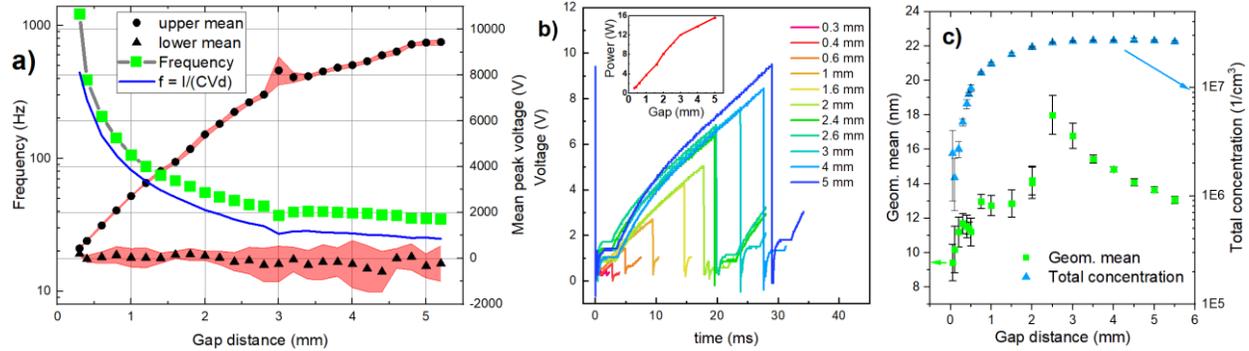

**Figure 4. a)** Effect of gap distance or electrode separation (*G*) on the frequency and upper and lower mean voltages for the sparks, whose bounds comprise the standard deviation of the upper and lower peak respectively. Conditions used were: *C*=10 nF, *V̇*=4 lpm, *I*=4 mA, *L*=100 µH, *Em*=Fe. **b)** Shape of a single spark and duration between representative sparks produced with the same conditions as in a). The inset shows the power released in the sparks as a function of gap conditions following Eq. 2 ( $E=0.5CV_d^2$ ). **c)** Total concentration and geometric mean particle size using conditions: *C*=2 nF, *V̇*=3 lpm N₂, *I*=1 mA, *L*=1 µH, *Em*=Fe. The results are in qualitative agreement with [52]. The size distributions can be found in SI (Fig. S6).

The effect of electrode separation on $V_d$, power, *f*, particle size, and concentration are summarized in Fig. 4. Fig. 4a and b show that as the gap between electrodes becomes larger, the upper mean of the voltage (*V*) increases, while *f* decreases (Eq. 5). Interestingly, the slope in the upper mean of $V_d$ changes at higher gap distances (*G*), suggesting two regimes. Below a *G*=3 mm, the $V_d$ rises more rapidly as a function of *G*. After this critical point, $V_d$ increases much more slowly as *G* increases. In the case of the lower mean, the variability becomes more prominent at larger *G*.

The electrode separation has a measurable effect on mean particle size and therefore, can be used to control it (Fig 4c), in agreement with previous studies [10],[21]. SEM/TEM images on the effect of *G* can be found in SI (Fig S3). A maximum in the mean particle size was found around *G*=2.5 mm (Fig. 4c) and seems to be linked to the change of regime in $V_d$. (Fig. 4a). The spark *f* is inversely proportional to *G*. But, the upper mean of $V_d$ increases at larger *G*, and the variability of the lower mean increases as well (Fig. 4a,b). The higher *G*, the more energy is dissipated per spark (inset of Fig. 4b), in agreement with [9] and [40].

The effect of *G* is interchangeable to the effect of pressure (*P*) and holds for different gases if normalized by their breakdown voltages ($V_b$) [37][61]. By following such approach, the results in Fig. 4 could be extrapolated to other gases and *P* by using Paschen´s law for discharge (Eq. 1) to obtain $V_b$ and then by using Eq 2 to obtain some insight on nanoparticle yield. Reports confirm this when comparing Ar vs N₂ [60]. $V_b$ for common gases grows as follows: He < Ar < H₂ < air < N₂ [18] [37] [55].

The slope of Fig. 4b indicates the charging rate of the capacitor at a value that varies from around 4 *10$^5$ V/s to 6*10$^5$ V/s using 10 nF and 4 mA, which is exactly what Eq. 5 predicts for the lower gaps, but this seems to deviate slightly at larger gaps. Generally, we observe a higher charging rate for larger *G*. Our observations are about 20 times higher than the dV/dt = 2*10$^4$ V/s reported for a spark using 2 nF and 0.04 mA [18].

The particle diameter reaching a maximum and then decreasing in Fig. 4c is somehow unexpected, yet it is in agreement with a previous report [52]. The resistance of the spark gap grows larger as a function of separation, and therefore larger *G*s require a higher $V_b$ to be overcome, producing more energetic sparks (inset of Fig. 4b). In agreement with [40]. Following such logic, one would expect a monotonic increase in particle size and concentration as a function of *G*. We observed a monotonic increase in concentration (although rather asymptotic above 2 mm) before the maximum at around 2.5 mm (Fig. 4c). A simple calculation also reveals a maximum of mass concentration at the same point. The same maximum was reported by Van Hoven et al. to occur around *G*=3 mm [52]. A discussion on a proposed explanation follows:

The presence of ions lowering $V_d$ at higher *G* seems to be experimentally invalidated as the $V_d$ increases continuously for larger gap distances (Fig. 4a,b). *f* gets larger for higher gaps, but this effect is insufficient to compensate for the $V_d^2$, which grows much more rapidly than the decrease from the frequency, as proven in the inset of Fig. 2b.

Van Hoven believed the effect was due to deviations in laminar flow [52]. Tabrizi et al. [18] did not see this maximum, but their reported experiments only covered up to 3 mm in Ar, which lies on the monotonic left side of our experimental results. Still, they observed a continuous decrease in the ratio of electrode mass loss/spark energy ratio, and explained this by assuming the spark energy becomes distributed over a larger volume, making the evaporation of the electrodes more difficult. Tabrizi et al. reported that only about 0.1% of the spark energy consumption is used for particle production [18]. This explanation seems feasible since Reinmann et al. reported that the current does not contribute to heating after 800 ns in their system [40].

We are inclined to believe that the spark becomes significantly thicker at *G*>2.5 mm, distributing its energy over a larger area and evaporating material less efficiently. Such effect could be described as an efficiency factor converting electric energy into material evaporation and nanoparticle mass production. The radius of the spark is taken into account in Eq. 13 to describe the threshold energy [24], so the efficiency can also be incorporated into it. The measurement of spark thickness is nevertheless challenging. Perhaps a good target for future research would be to observe how the ablated radius changes as a function of SDG parameters, particularly electrode separation.

## 2.4. Gas flow rate

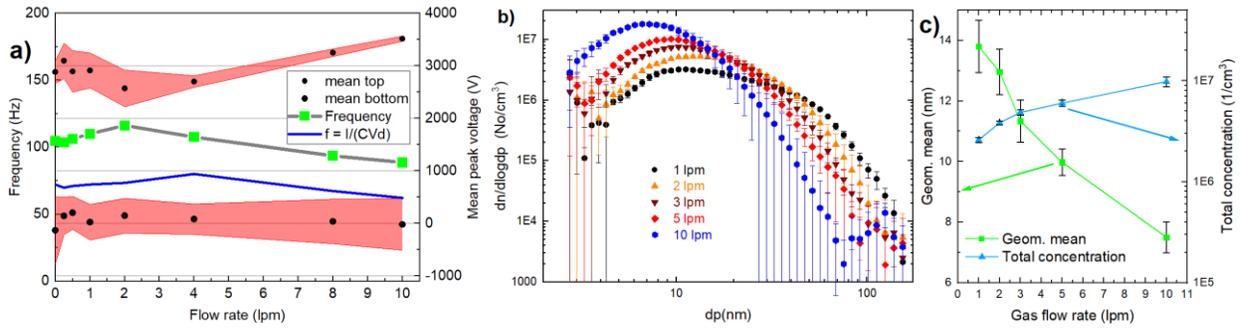

**Figure 5.** a) Effect of gas flow rate ($\dot{V}$) on frequency ($f$) and on discharge voltage ($V_d$) as measured at Node 3 ($N_3$) from Fig. 1. The figure shows a peak around 2 lpm but considering its value is only about 20% higher than the lowest part, it seems quite constant as compared to the effect of other variables on $f$. Conditions used were: $G$=0.5 mm, $C$=10 nF, $I$=4 mA, $L$=100 µH, $E_m$=Fe. b) Particle size distribution as measured after the 2 m SS sampling pipe. c) Total concentration and geometric mean particle size. Conditions used for b) and c) were: $G$=1 mm, $C$=2 nF, $I$=1 mA, $L$=1 µH, $E_m$=Fe.

This section presents the effects on the system variables when the flow rate ($\dot{V}$) is varied in our experimental chamber. Flow dynamics may be different for other geometries [62]. The frequency ($f$) increases continuously as a function of $\dot{V}$ up to 2 lpm and then decreases as $\dot{V}$ increases further (Fig. 5a). Nevertheless, the dependence of $f$ is very modest on $\dot{V}$ compared to its dependence due to current, or gap distance. $V_d$ first decreases slightly up to around 4 lpm and then increases linearly for higher $\dot{V}$, matching the $f$ trend around 2-4 lpm. Below this range, there seems to be insufficient flushing for the ions between the spark, so the breakdown voltage oscillates significantly. Above this flow range, the reproducibility of $V_d$ in the spark increases substantially with the standard deviation bounds of $V_d$ becoming narrower at higher $\dot{V}$ (Fig. 5a). This suggests better spark-to-spark consistency, achieved through better removal of ions and particles between the electrodes at higher $\dot{V}$.

A similar trend can be observed for the slope of the change in dp of the agglomerates. By increasing the flow rate of the carrier gas, a higher concentration of aerosol agglomerates of ever-decreasing diameter and narrower size distributions are obtained (Fig. 5b,c), in agreement with reports ranging $\dot{V}$ from 0.4 – 2 lpm [52], 0.5 – 3.3 lpm [25], 1 – 5 lpm. [44], 1 – 9 lpm [60], 2.9 – 3.7 [63], 3 – 6 lpm [12], 3 – 15 lpm [21]. In our case, the maximum mean particle size was observed at the lowest $\dot{V}$ tested (1 lpm). But a report shows the maximum does not occur at a negligible flow, but around 1.7 lpm [60], such value is near our peak in frequency around 2 lpm (Fig. 5a).

Aerosol distribution statistics show a significant effect when variations in flow conditions affect the coagulation rate [63]. For SDG, growth is dominated by two regimes. Between the electrodes, the flow is laminar and diffusion dominates. After the electrodes, it becomes turbulent and mixing dominates. Laminar flows can produce a higher yield of particles and narrower distributions [12]. Non-agglomerated particles can be obtained if the gas flow rate is sufficient to completely flush $G$ before a new spark occurs, as given by $f$ [10]. In such conditions, the number concentration of particles is described by Eq. 6 [12]. Experience says that by adjusting $\dot{V}, I, f, C$ and recharge voltages, the geometric mean can be tuned from 3 nm upwards [48]. The mean size of our Fe particles generated at the maximum flow rate tested of 10 lpm is about 7 nm (Fig. 5c). The total aerosol mass produced as a function of flow rate reportedly increases for larger flows up to around 1 lpm and then following a plateau afterwards [52]. We do not see the increasing regime as our experimental results start at 1 lpm (minimum flow for the SMPS) but we do see a plateau of mass concentration up to 10 lpm.

## 2.5. Resistance

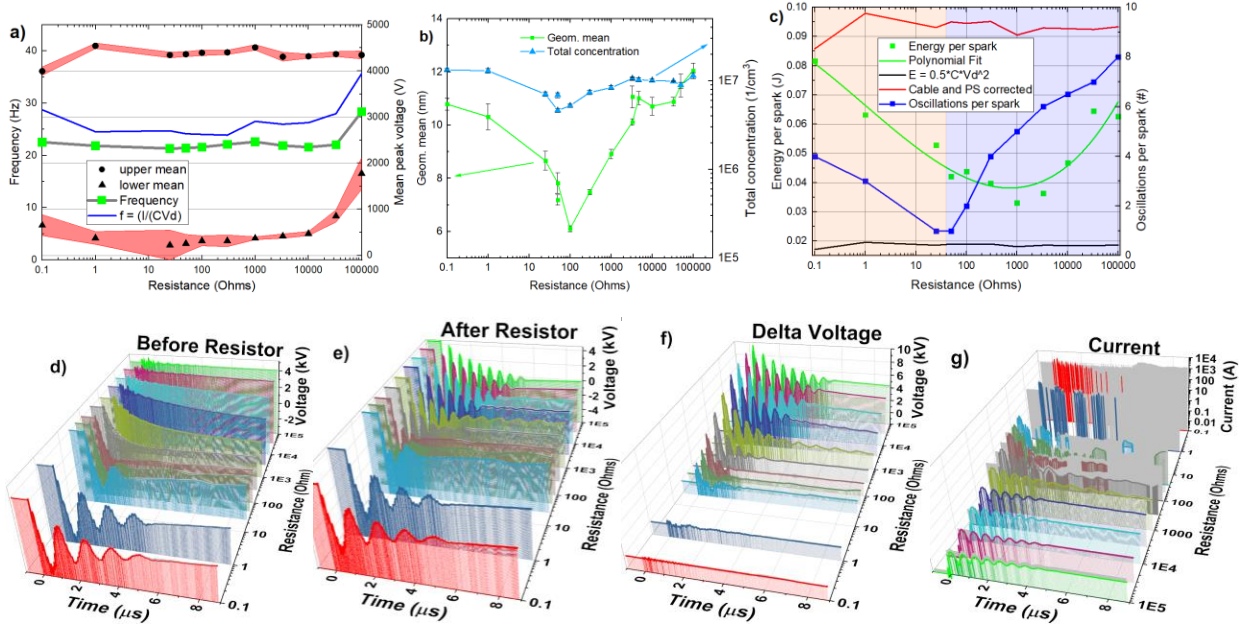

**Figure 6. a)** Effect of resistance R1 after the inductor as measured in Node 3 ($N_3$) of Figure 1 on the breakdown voltage and on the frequency effect of the resistance R1 after the inductor corrected with 7nF of HV source and circuit. **b)** Geometric mean particle size and total concentration of the particles produced at different resistances. The conditions used for SMPS data were: **G**=1 mm ,**C**=2 nF, **V̇**=3 l pm $N_2$, **I**=1 mA, **L**=1 µH, **Em**=Fe. **c)** Calculation of energy per spark based on measured voltage and calculated current, and number of oscillations per spark. The orange background color means the system is underdamped and the blue background range indicates the circuit is overdamped. Based on Q or on the damping criteria. **d)** V before **R1** at $N_2$ (Figure 1). **e)** V after **R1** at $N_3$. **f)** ΔV calculated. **g)** I calculated based on Ohms law (a 2D plot for the same data is available in Fig. S8). Note the change in resistance direction of the g) plot for better visualization. For estimating energy at resistances higher than 3300 Ω, the µs resolved plots did not plateau and spark voltage ranging ms was used (Figure S9). Conditions used for all cases: **G**=1 mm ,**C**=2 nF, **V̇**=3 lpm $N_2$, **I**=1 mA, **L**=1 µH, **Em**=Fe.

Fig. 6a shows that resistance (**R**) has a negligible effect on the discharge voltage (**$V_d$**) (upper and lower mean of the voltage) and the frequency (**f**) of the spark, except at 100 kΩ, where **f** increases about 20 % because the **V** lower mean peak does not go below 1kV, allowing for a faster capacitor recharge (Fig. S10). Due to the rather low current and capacitance used, the lower mean of the breakdown voltage never reaches zero (Fig. S9). This is taken into consideration for calculating **$V_d$**. Based on Eq. 5, **f** should not depend much on resistance unless it changes **$V_d$**, in agreement with Fig. 6a.

One would expect that adding a resistance before the spark would cause energy loss and therefore have a monotonic decrease in nanoparticle concentration and size. However, both the aerosol geometric mean size and the total concentration have a minimum at around 100 Ω, and then rise again (Fig. 6b).

The voltage before the spark (**$N_2$**) was measured at different **R** and is plotted as a function of time for the first 9 µs after the spark onset (Fig. 6d). Results show that there are two regimes regarding voltage oscillation behavior. For **R**<1000 Ω, the oscillations number increases for smaller resistances. For **R**> 1000 Ω, the system does not oscillate and is overdamped. The higher the **R**, the smaller the reaction of the voltage before the resistor (**R1**) to the drop caused by the spark, barely decreasing 9 μs after the spark onset at the 100 kΩ case and taking in the order of ms to reach the minimum (Fig S10). Based on this and the results of the inductor section, a schematic mechanistic explanation is provided in Fig. S7.

The oscillations in the spark (measured at **N₃** after **R1**) appear to have two regimes (Fig. 6e). In the first one, at low **R**s, the oscillation has a larger amplitude. At **R1** = 0.1 Ω, **V** oscillates 4 times (1 oscillation = 1 full up and down cycle) and as **R** increased, the oscillation decreased. Then, at **R1** slightly below 100 Ω, the voltage before **R1** becomes critically damped and no longer oscillates. In the second regime, at **R**>1000 Ω, oscillation occurs again but with a different **f** and the larger the **R**, the more oscillations it produces in the spark. A summary of the number of oscillations in each condition is given in Fig 6c. The 1000 Ω case shows the longest ripples after **R1** once the spark oscillation is over (Fig 6e). This might be caused by a resonant RLC interaction between both sides of the circuit.

The resonant frequency (**f₀**) for the whole circuit, including the additional circuit **C**= 7+1+2 nF = 10 nF. For the 8 m cable **C**=1+3 nF. **f₀** for the whole system as calculated based on Eq. 19 is 0.783 MHz. In the cable case, it is 1 nF and 3 μH so the resonant frequency is estimated to be: 2.8 MHz. These calculations compare well with the measured **f** in both spark regimes. The oscillations in the regime of low **R** (<1000 Ω) are 750 kHz. The ones at high **R** are 1.5 MHz. **f** might be overestimated in the cable case due to additional components not being considered between **R1** and the HV cable, such as connectors and electrodes. The calculation of dampening for the whole circuit predicts that the system is underdamped at low **R**. At high **R** it becomes overdamped, in agreement with Fig 6d. The critically damped condition was calculated to occur around 50 Ω. This underestimation might be caused by not considering the RLC properties of the electrodes.

Time-resolved measurement of the voltage before (Fig. 6d) and after (Fig. 6e) **R1** provides insight into the spark behavior and allows the calculation of time-resolved delta voltage, **ΔV** (Fig 6f). Based on Ohm´s law, one can estimate current (**I**) (Fig 6g), power and by integrating it, the energy per spark (Fig. 6c). 2D plots can be found in Fig. S8-S10. The equations for these calculations follow:

$$I = \frac{V}{R} \qquad \text{Eq 15}$$

, where **I** is the current, and **V** the delta voltage across a resistance **R**. Power is given by:

$$P = \frac{Vq}{t} = VI \qquad \text{Eq 16}$$

, where **q** is electric charge in Coulombs and **t** is time in s. It follows that energy is given by:

$$E = Pt = VIt \qquad \text{Eq 17}$$

For transitory or non constant processes, where **V** and **I** are changing as a function of time, such as in a spark, the energy can be calculated by:

$$E = \int_0^{end} V_t I_t dt \qquad \text{Eq 18}$$

To better represent the system and minimize SDG intrinsic spark-to-spark breakdown voltage discrepancy [15],[18], 32 triggered measurements were averaged to construct Fig. 6d and 6e. **ΔV** (Fig 6f) is obtained by subtracting **V_{before}** (Fig. 6d) - **V_{after}** (Fig. 6e). Time-resolved current (Fig. 6g) is obtained from Eq. 15. The current calculation is susceptible to significant magnification by instrumental errors due to a division by a very small resistance (see Materials and methods section). Integration of amplified error in current to estimate energy would be misleading. Therefore, specific temporal values of current are not included in the integral when the standard deviation is more than twice its mean value. The origin of this measurement artifact is the small but constant *ΔV* remaining after the spark ripple voltage has come back close to zero, which is later magnified by dividing by a small *R* (Eq. 15). This precaution is only taken up to 1000 Ω and is unnecessary for larger *R*s (Fig. S9 and S10). The current achieved during discharge reaches over 200 A but only for about 5 μs. This behavior has also been observed for Cu, Pd and Pt.[14]

Eq. 2 predicts energy to be independent of *R*. In the measurements of this section, *C* was constant and the $V_d$ does not change enough (Fig. 6a) to explain the minimum in the energy per spark depicted in Fig. 6c. However, it is consistent with the measured mean size and concentration of the aerosol generated. Therefore, an additional effect correlated to the RLC circuit nature must be taking a role and we believe it is related to the *Q* factor and the dampening criteria. A discussion on these lines follows.

**RLC and LC circuits**

The resonant frequency of an RLC circuit is an important parameter and has a direct impact on SDG behavior. Our circuit might look like a parallel LC circuit. However, this is only the case during charge. For discharge, it is actually in series. The topology of the circuit with respect to current for charge and discharge is sketched in Fig. S7. The resonance frequency is an important parameter. It is given for both LC and RLC circuits as:

$$f_0 = \frac{w_0}{2\pi} = \frac{1}{2\pi\sqrt{LC}} \qquad \text{Eq 19}$$

, where $w_0$ is the angular frequency and $f_0$ is the frequency in Hz. For series circuits, if *f* < $f_0$, the circuit is capacitive. Otherwise, if *f* > $f_0$, the circuit is inductive. For parallel circuits, the criteria are inverse. But in our case, the parallel circuit is not closed as the spark is absent during charge. We cannot modify the natural frequency of the spark oscillation and normally the cable length is also constant regardless of the experiment. For this reason, to modify the conditions of the resonance frequency one modifies the RLC components directly. In this section, we modified the resistance component only. The resistance increases the decay rate of the oscillations, known as damping.

The quality factor (*Q*) is a widespread parameter used in resonance. For a series resonant RLC circuit, *Q* indicates how underdamped an oscillator or resonator is. A high Q means the oscillation lasts long, and a low *Q* means the oscillation dies out quickly. *Q* for a series RLC circuit is given by:

$$Q = \frac{1}{\omega_0 RC} = \frac{1}{R}\sqrt{\frac{L}{C}} \qquad \text{Eq 20}$$

, where *R* is resistance, *L* is inductance and *C* is capacitance. The oscillation indicates how often the polarity is reversed in the electrodes, and therefore, that both electrodes are being ablated [10]. In unipolar discharge, positive ions ablate more strongly the anode [10]. The damped oscillation is due to the inductive component of the circuit [18]. *Q* is critically damped at 0.5. Below this value, the system is overdamped and above it is underdamped. Higher Q values oscillate for more cycles than smaller *Q* values. *Q* for a parallel resonant RLC is the reciprocal of Eq. 20:

$$Q = R\sqrt{\frac{C}{L}} \qquad \text{Eq 21}$$

For circuits where **L** and **C** are in parallel, the resonance frequency occurs at a maximum impedance rather than at a minimum, but it is still called a resonance frequency. These circuits are also referred to as antiresonators. The definition of damped and underdamped is inverse for series and parallel circuits, creating confusion. In our case, the value of **Q** that is critically damped matches with the maximum in energy loss (Fig. 6c). The critically damped response is the fastest possible circuit response that does not oscillate. **Q** is the inverse of fractional bandwidth $B_f$, where $B_f$ is:

$$B_f = \frac{\Delta\omega}{\omega_0} \qquad \text{Eq 22}$$

$$\Delta\omega = \omega_2 - \omega_1 = 2\alpha \qquad \text{Eq 23}$$

, where $\Delta\omega$ is the bandwidth, $\omega_1$ is the lower half power frequency, and $\omega_2$ is the upper half power frequency. Underdamped conditions such as low resistance allow the oscillation before and after the resistor to pass almost unaffected, whereas damped conditions of high resistance have oscillations after **R1**, but before **R1** the circuit needs time to react. On the other hand, precisely in the critically damped system, the circuit acts as a bandpass filter at the natural resonance frequency of the spark, and therefore, absorbs energy from the circuit preventing its utilization in the spark, having as a consequence, lower energy with its corresponding smaller particles and lower total concentrations. The gap can be modeled as a purely resistive component without any inductive component. With this, a second-order differential equation is obtained [14].

$$L\frac{d^2Q}{dt^2} + R\frac{dQ}{dt} + \frac{Q}{C} = 0 \qquad \text{Eq 24}$$

, where **Q** is the charge on the capacitor, **L** is the inductance of the cable used and **C** the capacitance. Similar to **Q**, the following damping criteria can be established:

$$R^2 - 4\frac{L}{C} < 0 \text{ underdamped system} \qquad \text{Eq 25}$$

$$R^2 - 4\frac{L}{C} = 0 \text{ critically damped system} \qquad \text{Eq 26}$$

$$R^2 - 4\frac{L}{C} > 0 \text{ overdamped system} \qquad \text{Eq 27}$$

Based on such criteria, considering the prediction from the RLC system using the correction for the cable and circuit capacitance and inductance, the critically damped system should be between 25 and 50 Ohms. This is consistent with the experiments and also near to the minimum observed in terms of geometric mean particle size, total concentration and energy calculated per spark. This suggests a direct link between the characteristics of the aerosols and the RLC parameters of the circuit. A comparable analysis can be achieved by means of the **Q** factor. The damping criteria are equivalent to **Q** but the numerical values for a critically damped system are different; for **Q** it is at 0.5, while the damping criteria dictate zero. **Q** has an asymptote at zero and therefore, we used the dampening criteria for this study.

**Spark duration**

One might be tempted to say that the spark lifetime is given by the time at which **ΔV** (**V** measured at **N₃** minus the one at **N₂**) is substantial (in the kV range). This ranges between ~5 μs for the 0.1 Ω case (Fig. 6e) to ~2 ms for the 100 kΩ case (see Fig. S10 for ms resolution). However, the oscillation within the spark measured at **N₃** is within the range of 2 – 5μs for all **R**s tested from 0.1 to 100 000 Ω. This observation supports that the end of the spark occurs when the voltage of the spark side is lower than kVs (at **N₃**). The spark duration as measured at **N₃** goes from 5 μs at 0.1 Ω down to 2 μs near critical conditions (25-100 Ω) and then up again to 5 μs at higher resistances (Fig. 6e). The observed ΔV lasts longer for larger resistances since they require some time to reach equal voltage on both ends. The whole system might be behaving as a bandstop filter at the critical conditions, preventing the circuit side of the spark from contributing to define the properties of the spark when the circuit is overdamped.

**Spark energy and oscillations**

Fig. 6c shows a minimum in energy. This is surprising, as one would expect that larger **R**s before the spark would cause a monotonic power loss with a weakening of the spark energy. As predicted in Fig. 2a, mean size and total number concentration measurements (Fig. 6b) also support this. Stability criteria, oscillations per spark, mean size and number concentration all match pretty well between 25 – 100 Ω. On the other hand, energy per spark also predicts a minimum, but at around 1000 Ω. The slight discrepancy in the predictions from different methods could be attributed to intrinsic inaccuracies of each method and non-idealities in the system and measuring devices. The system is underdamped at **R** < ~25 - 50 Ω and overdamped at **R** above this value. This seems to affect the energy per spark, which ultimately defines mean particle size and concentration [10].

The efficiency of the transmission of energy from the circuit into the spark seems to increase at higher **R**s. This is supported not only by energy calculations per spark, but also by the generation of nanoparticles in terms of total concentration and particle size that are also continuously increasing as compared to their minimum at 100 Ω. When looking at the times the oscillation occurs in Fig. 6 d,e it seems like the number of oscillations is perhaps increasing the efficiency with which the energy in the circuit is converted into nanoparticles.

When the damping criteria are < 0, the RLC system is underdamped, so the voltage demanded by the spark is delivered by the cable combined with the RLC circuit. On the other hand, when the damping criteria are > 0, it means the response of the circuit is too slow before **R1** to affect the spark directly, and the spark features resemble the limitations of RLC given by the cable after **R1**. For this reason, an increase in **f** is observed in this second regime as compared to the first one. The oscillation period (time taken to complete one cycle of oscillation) for the whole circuit is around 1.3 μs. For the cable and spark section, it is around 0.67 μs. At the critical condition, the cable oscillation **f** and the circuit **f** seem to dampen one another and cause the oscillation to die out more quickly, releasing less energy in each spark and therefore producing fewer and smaller aerosols. This compares well to the oscillation period reported by Tabrizi et al. with a period of 0.38 μs. However, our spark oscillates at a lower rate and also tends to last at least 5 μs, whereas the one of Tabrizi et al. lasts only 2 μs. They used **C** = 2 nF, **L** = 1.9 μH and 5.1 Ω, but did not report their cable length [18]. The inductance of aluminum finned resistors as the ones used here, should be negligible (0.02 – 2 μH).

## 2.6. Capacitance

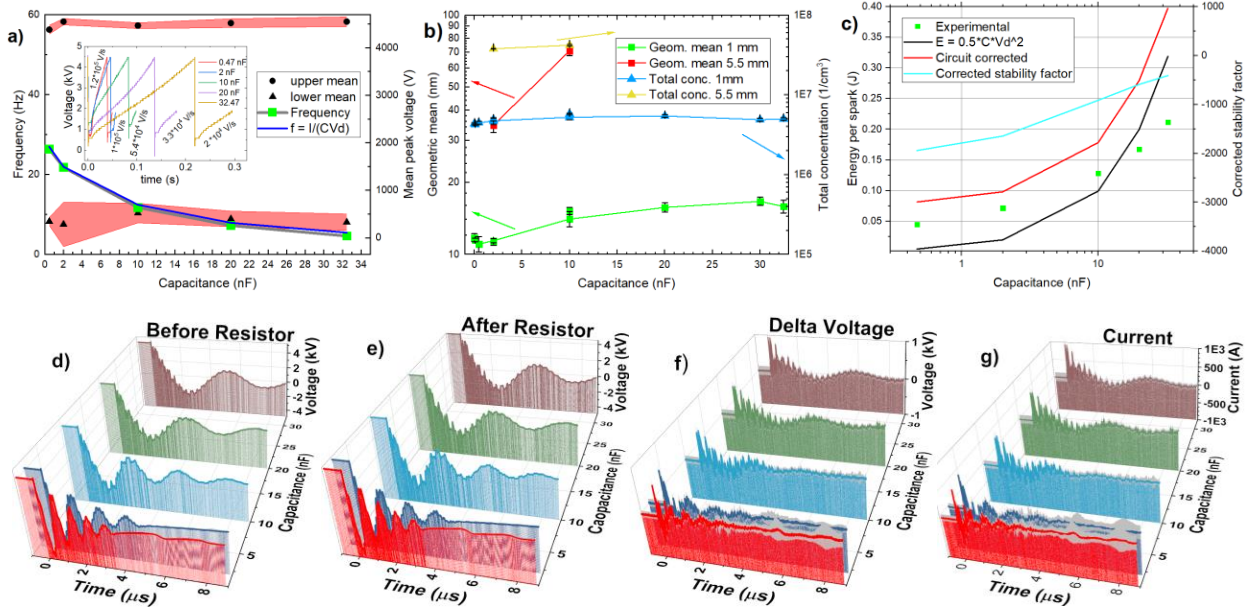

**Figure 7. a)** Effect of capacitance (**C**) on discharge voltage (**V_d**), lower voltage and frequency (**f**) using spark. Inset shows full discharge and charge cycle of sparking as a function of different **C**. The conditions used were: **G**=1 mm, **V̇**=3 lpm $N_2$, **I**=1 mA, **L**=1 µH, **Em**=Fe. **b)** Geometric mean particle size and total concentration of nanoparticles produced using **G**=1 or 5.5 mm, **V̇**=3 lpm $N_2$, **I**=10 mA, **L**=1 µH, **Em**=Fe. **c)** Estimated energy from experiments plotted vs the nominal **C**, comparison vs Eq. 2, with and without correction from cable and circuit **C** values. The trend of the predictions is correct when compared to experimental measurements, but the quantitative values are overpredicted by up to a factor of 2. On the right axis the stability factor introduced in Eq. 25, $R^2 - 4\frac{L}{C} < 0$; the factor is negative so the system is underdamped for all capacitances tested. **d)** Voltage before a 1 Ω **R1** resistor measured at $N_2$ node, **e)** after the resistor at Node $N_3$. **f)** ΔV by subtracting e from d. **g)** Estimation of current through spark based on Ohm´s law.

The sparking frequency (**f**) decreases slightly at higher capacitances (**C**) (Fig. 7a). On the other hand, the lower and upper voltage means remain virtually constant. In general, spark reproducibility worsens at higher **C,** as the increasing dispersion in the standard deviation bound indicates. The inset of Fig. 7a shows the measured charging rate for different **C**s, ranging between ~$1.2*10^5$ V/s for 0.47 nF, and $2*10^4$ V/s for 32.47 nF (conditions were: **G**=1mm, **V̇**=3 lpm $N_2$, **I**=1mA, **L**=1µH, **Em**=Fe). The charging rate is slower for higher **C**s as more energy and time is required to increase the voltage. Therefore, sparks occur at a lower **f**, as shown in Fig. 7a (all other variables remaining constant). This compares well to the $3.5*10^6$ V/s as for the 1 kHz, 1.2 kV and 0.5 mm reported by Pfeiffer et al. [15], the dV/dt = $2*10^4$ V/s for a C=2 nF and I=0.04 mA [18]. Or the $1*10^6$ V/s and $3.5*10^6$ V/s applying for glow regime and capacitive discharge in He [15].

The effect of **C** in particle size distribution is presented in Fig. 7b. The agglomerate geometric mean size, as measured via SMPS, gets larger for higher **C**. The same trend is seen in TEM images (Fig. S3). These observations are in agreement with previous studies [10][18][44]. The same trend has been reported in terms of nanoparticle concentration, where an increase from $1.38*10^{-4}$ to $2.19*10^{-4}$ kg/ $m^3$ was achieved by increasing **C** from 20 to 100 nF [44]. Horvath et al. [17] observed that the size distribution narrowed at higher charges in their capacitor (6 µF). We did not see this trend, but we only tested our SDG up to 32.47 nF.

The energy per spark was estimated using Ohm´s law approach from time-resolved voltage characterizations. The results are presented in Fig. 7c and are compared to calculations from Eq. 2 (correlation between energy, $C$ and $V_d$) using nominal and circuit and cable corrected $C$ values. The results are in agreement with Garwin et al. [9], who mention that both $C$ and $V_b$ given by gap spacing, pressure, and gas used increase energy dissipation [9]. Horvath and Gangl reported a non-linear correlation between capacitor´s energy and aerosol mass produced. Such dependence plateaus upon capacitor saturation as shown gravimetrically and by DMA [17]. Tabrizi et al. [18] reported a direct proportionality between capacitance and spark energy. They observed via TEM and DMA that more mass is vaporized at higher capacitances, causing the mean size and particle concentration to increase for both agglomerate and primary particle [44]. Tabrizi et al. [18] reported their maximum spark energy at 250 mJ. $E$ and a charge $CV_d$ of $6.25*10^{14}$ electrons/spark [18].

$C$ determines the spark energy and its duration. Our $C$ results seem to agree with Reinmann et al. [40], who reported that the current does not contribute to heating after 800 ns in their system. In our case, the peak lasted for about 4 µs. The difference might well be due to the different circuit conditions used. We observed that increasing $C$ increased spark energy. The corrected stability factor shows that the system is in all cases underdamped (Fig. 7c, right axis). The only way to use these ranges of $C$ and inductance ($L$) to achieve a critically damped system is by using a higher resistance, as in the previous section.

Meuller et al. [12] report that the effect of capacitance ($C$) is on determining the spark energy, whereas $L$ affects spark duration. They say that if the spark duration is longer than the duration of the electrode heating, more material can evaporate [12]. We agree that higher $C$s produce more energetic sparks but we observe that higher $C$s also cause longer sparks (Fig. 7d). Our used $C$ ranges from 0.1 to 32.47 nF, which is close to the range of many studies: 100 pF – 1 nF [25], 6nF [40], 0.47 – 47 nF [48], 45 nF [15]. Perhaps the largest capacitance reported for an SDG is the 6 µF by Horvath et al. [17]. Our results are in agreement with previous reports saying that for a given RLC circuit, at constant $L$, an increase in $C$ results in a decrease in $f$, effectively widening the oscillations of the spark [64]. The period of the spark is the following: for 0.47 nF: 0.8 µs, for 2 nF: 1.33 µs, for 10 nF: 2.67 µs, for 20 nF: 4 µs and for 32.47 nF: 5 µs. The period of the spark increases proportionally with $C$.

## 2.7. Inductance

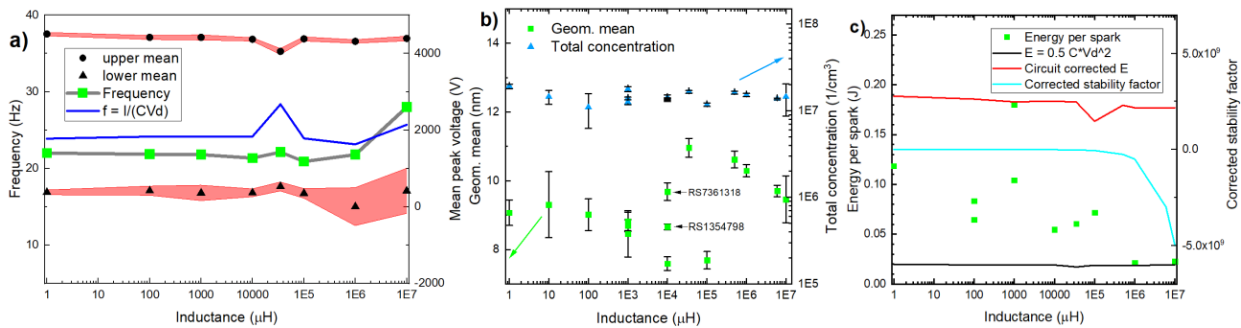

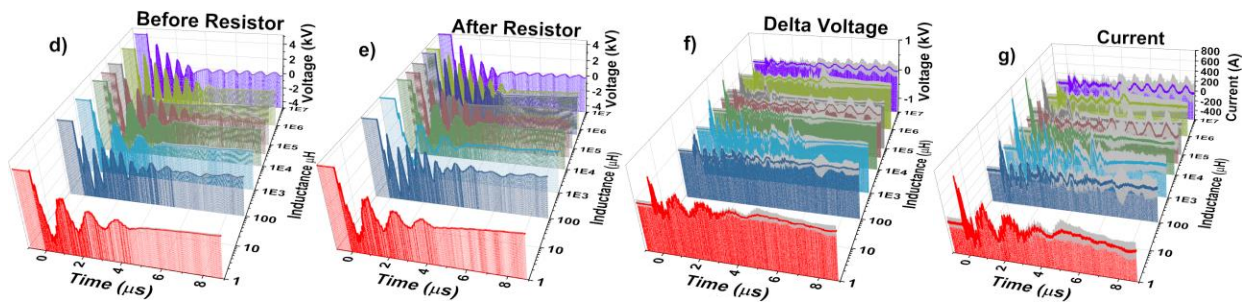

**Figure 8. a)** Effect of inductance ($L$) on frequency ($f$), discharge voltage ($V_d$) and minimum voltage after the spark. All three variables remain fairly constant throughout the whole range of $L$ tested. Conditions used were: $G$=1 mm, $C$=2 nF, $\dot{V}$=3 lpm N$_2$, $I$=1 mA, $E_m$=Fe. **b)** Geometric mean particle size and total concentration of aerosol produced remains fairly constant for the whole experimental scope, ranging 9 +/- 2 nm for all cases. The full particle size distribution can be found in SI (Fig. S14a). Conditions used were the same as in Fig. a. **c)** The corrected stability factor is negative for the whole regime and becomes more dominantly negative at higher $L$s. Since the stability factor does not cross towards positive numbers, it remains in the underdamped regime for all measurements (see Eq. 25). **d)** Mean voltage of the spark before an $R1$ of 1 Ω. **e)** after resistor $R1$ at node $N_3$, **f)** delta voltage, **g)** current calculated via Ohm´s law.

In general, inductance ($L$) is a variable whose effect is more difficult to observe, due to the intrinsic difficulties of implementing commercial inductors at the high values of $f$, $I$ and $V$ typical in an SDG. Results show that $L$ has a small effect on the upper and lower mean of $V_d$, and $f$ is only responding to the changes in $V_d$ (Fig. 8a). The geometric mean nanoparticle size for all $L$ cases tested ranges between 7 and 11 nm, without a very clear trend, but 9 +/- 2 nm is sufficiently constant to infer that there is a negligible effect in the ranges of $L$ cases tested for the production of nanoparticles. The total concentration appears to be constant too (Fig. 8b). The stability factor is in all cases <0, which means that the system is underdamped and oscillates for all cases (Fig. 8c). Also, some estimations of energy per spark based on Eq. 18, with a correction to include the circuit influence and uncorrected (taking only nominal values for RLC components), are compared vs estimations of energy per spark obtained via Ohm´s law using Fig. 8d,e as a basis. $V$ measured before a $R1$ of 1 Ω in N$_2$ (Fig. 8d) and after $R1$, at node $N_3$ (Fig. 8e) were acquired for nominal inductors from 1 μH to 10 H. $\Delta V$ (Fig. 8f) is given by subtracting Fig. 8d from Fig 8e and current (Fig. 8g) is calculated via Ohms law as previously described.

All real inductors have a finite value of $R$. Unfortunately, this value is not constant for inductors with different $L$. In addition, the inductors get saturated at currents much smaller than the typical currents in SDGs. These features make it difficult to study the isolated effect of inductance in the spark. This causes some irregularities in the trend of Fig 8 d,e. However, by taking into consideration both the saturation current and $L$, the trend becomes clearer. The spark is considered to be over after the large rippling in the spark reaches a value lower than about 500 V from the settled point. After the spark is over, especially for the 35 mH and 10 H cases which oscillate the most with ripples that last up to 50 ms for the 10 H case (Fig. S16). Since both have the largest value of inductance multiplied by saturation current, both seem to be damped after passing through the resistor. The frequency of the ripples after the spark seems to be the same as during the spark, suggesting that it is the circuit/cable itself the one which determines the spark properties and not the spark itself. A puzzling case is the one at 100 μH, which seems to oscillate erratically, perhaps because its resistance is very low, and its inductance is just the right value to cause the oscillation from both sides of the circuit to interfere with one another. This case also has the largest error bar for the particle size, which represents the standard deviation of the mean size for several measurements. Spark shape reproducibility was confirmed by using different time acquisition ranges and all ripples match pretty well (Fig. S13).

The maximum particle size is achieved at 35mH, which is the inductor of high **L** with the lowest resistance (0.79 Ω). This inductor also causes the maximum spark oscillations (excluding plots that appear to be of mixed origin). Then, particles decrease in size for increasing **L**, but the resistance of the inductor also increases, so the culprit for the decrease in size is considered to be the resistance itself. The fact that inductors have a resistance is a known problem [64], which causes significant, often undesirable effects on the circuit. It seems both resistance and inductance block the first RLC circuit, and the observed oscillation is attributable to the cable, which might also behave as an RLC circuit.

Inductors have an important impact on the electrical characteristics measurable at $N_2$ and $N_3$. Very high inductance causes ripples at 10 H. The shape of the spark seems to change but the geometric mean of the aerosol seems to have a minimum at around 10 - 100 mH. This is nevertheless a relatively small effect (if any) compared to other variables. It might be a consequence of a short circuit within the inductors and their intrinsic **R** values.

The 10 H case is the one with the largest **R** and **L**. The high **R** is the reason why the mean oscillations die out fairly quickly, and the high **L** might be the reason why the small ripples remain for much longer. The one that looks like a critical condition at 100 000 µH is much rougher and has more intermediate peaks than the ones observed at the critical condition of 50 Ω in the resistance characterization, which is much smoother. The inductance has an effect even at the ms regime, which is not seen at other conditions (compare Fig. S16 vs S10). There is an interesting effect for the cases of high inductance. The last 4 cases (100 mH – 10 H) show, after about 1 ms, an abysm or peak that is quite reproducible. This could be due to the current reaching levels lower than saturation. Therefore, the whole strength of the inductance starts finally taking a role (See Fig. S16).

The saturation current of all inductors is much lower than the current observed during the spark. The effect of the inductors measured here seems to be more influenced by their **R** than by their **L**. The inductor gets saturated, its inductance drops and the resistive section causes the effective emergence of 3 RLC circuits in series. The 100 µH inductor, shows irregular oscillations which appear to be caused by substantial interference between two RLC circuits. This can be explained by its resistance of 0.24 Ω, which is close to the 1 Ω used between the measurement nodes **$N_2$** and **$N_3$**.

The presence of inductance combined with resistance and, at the same time, having a saturation current much lower than the spark current makes the analysis cumbersome and non-trivial. For future analysis, we recommend utilizing inductors with resistances as close as possible to one another and with a saturation current in excess of 200 A. To our knowledge, such inductors might be challenging to acquire in the current market.

**Characteristic times**

Data suggest that the higher inductances need more time to reach their final energy position. High inductances decrease the reproducibility of the system and sparks. We believe the energy per spark calculation suffers more noise than in the calculations for **R** and **C** in previous sections. The high inductance cases seem to be underestimating the energy, as the oscillating nature of the high inductances brings the mean close to zero, but the error bars show such oscillating nature (this is particularly notorious in the

10 H and 35mH cases in Fig. 8g). An additional oscillation is observed for the high inductances, which begins at about 100 µs and reaches beyond 4 ms (Fig. S16).

Among all SDG variables, perhaps inductance is the one that has been studied the least [12]. In the study of Tabrizi et al. [18], no inductor was used but an equivalent drawing shows *L* as the inductor of the cable. The system oscillates with a period of roughly ***T = 2 π(LC)$^{0.5}$***, so ***L*** = 1.9 µH. The damping time constant of the curve is ***τ***=0.72 µs. Then in the formula of ***τ = 2L/R*** the resistance is 5.1 Ω. A weak resistance is needed for the weak dampening assumption ***R< 2(L/C)$^{0.5}$*** .[18]. Our period is of 0.67 µs for the cable-dominated regime (underdamped) and of 1.33 µs for the circuit-dominated regime (overdamped). Based on these calculations, for the underdamped case, an after-inductor circuit inductance of 11 µH is obtained, the cable contributing 4 µH. The calculated resistance of the spark is 34 Ω. For the overdamped conditions, the inductance for the whole circuit is calculated to be 40 µH and the spark resistance 7Ω.

**Stability criteria**

At all conditions tested for *L*, the system was underdamped and seems to be inductance dominated. Based on *Q* factor, the criterion is always too high in *Q* and does not migrate from very high to very low *Q* values. The stability parameter is underdamped across the 7 orders of magnitude of inductance tested. No significant effect was measurable in part because the inductance of the cable was already above the critical threshold. Testing with a shorter cable could provide more insight into the overdamped transition based on *L*. At high *Q*s, the system is underdamped and behaves as resistive/capacitive. *L* definitely has an effect on the SDG but it is not easy to observe since each inductor had a different resistance. This is a well-known problem in electronics. It was not possible for us to acquire inductors with the same resistance for the 7 orders of magnitude tested, and it is the cause of the non-smooth transition between the low and high inductances tested. The specs of the inductors can be found in Table S2. There seems to be a direct correlation between the number of oscillations and the particle size. The largest particles were produced at conditions of 35 mH, where the number of oscillations was maximum and without destructive interference.

The inductance in the circuit seems responsible for increasing the length of an individual sparking event, but it is counteracted by the effect of resistance, which shortens the spark oscillation to settle more rapidly, as seen in the resistance effect section. A larger inductance causes the spark to last longer. This can be seen by taking a close look at Fig. S15. This observation seems to be in agreement with Meuller et al. [12] about inductance having an effect on the spark duration and also with the report that the oscillations in experimental waveforms have been attributed to the inductance, capacitance of the measuring probes and their interactions with the spark, electrode and cable [40].

During discharge, it has been proposed that the circuit behaves as a purely resistive circuit with R$_{SG}$ and no significant inductive component [14]. This observation might emerge from most studies operating in conditions where the stability factor lies in the same regime. Even in our case, despite ranging 7 orders of magnitude in nominal inductance, all our experimentally tested inductance values were in the underdamped regime. The critical value for reaching critically damped conditions at the 2 nF and 1 Ω of the experimental data reported in Fig. 8 is of 0.5 nH or about 3 orders of magnitude lower than the lowest inductance achieved within the system of 4 µH, from which only the cable contributes with 3 µH. A way to achieve critical conditions without lowering inductance is by using a higher capacitance. However, not even using a 100 nF capacitor would be enough to observe the effect of inductance in a critically damped circuit, as this would still require an inductance of 0.1 µH, which is still over 30 times lower than the intrinsic *L* of the cable in our SDG. For this regime, the only way to achieve overdamped conditions seems to be by using a high *R*, or a shorter cable, hindering SDG movability.

The lowest concentration and smallest particles appear when the number of oscillations is the lowest (in agreement with the observations from the resistance section). This calls for searching conditions with the largest number of oscillations for maximizing particle mass production.

The shape of the sparks for the inductance results can be classified into 4 different groups:

1. Full interaction between the circuit and the cable towards the spark (schematically shown in Fig. S7). *L* values are 1 µH, 1 mH, and 10mH, with resistance values of 0.01, 1.4 and 39 Ω respectively. The *R* of the inductor is sufficiently small (<39 Ω), so the system behaves somehow similar to the underdamped conditions observed in the resistance section for cases of less than ~40 Ω (Fig. 6c).
2. The condition of 100 mH and 235 Ω looks like the critically damped case in the resistance section. Perhaps the inductive component requires a higher resistance to maintain critically damped conditions (compare with the critically damped condition of Fig. 6d. at 50 Ω and 1 µH). Also note that its saturation current is one of the lowest (0.02 A).
3. The 100 µH and 1 mH torus cases, present fairly high saturation currents of 1 A and 10 A. Such high saturation allows the inductive effect to become more important and, at the same time, the low resistances of 0.24 and 0.007 Ω cause the two signals to become mixed. 1 mH torus case is not plotted in Fig. 8 but can be seen in Fig. S15d.
4. The largest inductors case were the ones of 35 mH, 500 mH, 1 H, and 10 H, with resistances of 0.79, 30, 50, and 500 Ω, and saturation currents of 2, 0.3, 0.24, 0.2 and 0.05 A. This group behaves as a large resistance case from the resistance section. The inductor with the lowest resistance of this group (35 mH and 0.79 Ω), also has the highest saturation current, for this one, the inductance effect compensates the low resistance.

By using a metric to correct the effect of current saturation, a simplified approach is presented in the supplementary section, which is able to explain the spark shape as a function of the corrected inductance, calculated by multiplying the nominal resistance times saturation current and dividing by a hypothetical spark *I* of 200 A (Table S3). In our case it looks like the inductor has a certain resemblance to the resistance effect. It is a known problem that inductors come with a value of resistance and a saturation current that is often insufficient for an SDG. For this reason, observing the pure effect of induction remains challenging. Perhaps using a cable of different lengths would be the best way to characterize this.

## 3. Conclusion

In this work, we constructed a spark discharge generator (SDG) and summarized some important design considerations. We systematically observed the spark behavior and the nanoparticle agglomerates produced as a function of several variables: current, electrode gap distance, gas flow rate, resistance in the circuit, capacitance and inductance. We summarized results in terms of the electric behavior of the spark and the characteristics of the particles generated, all of them as a function of the different variables. The study allows pinpointing conditions required to increase SDG yield: the frequency between sparks needs to be increased as much as possible to improve the yield. Additionally, it seems productive to increase the oscillations between a single spark by choosing a large resistor near the spark followed by an inductor. A brief summary for each section follows:

**Design considerations:** the SDG should be constructed considering user and circuit safety. To protect the power supply without losing output, an optimal resistance value exists. The energy in the spark plays a critical role in the ablation of nanoparticles from the electrodes. The SDG produces a large fraction of charged particles, and in some circumstances (normally undesirable), particles of several μm can be ejected from the electrodes.

**Current:** for dynamic generator conditions where the aerosol formed is at a concentration above ~$10^8$ part/$cm^3$, the particles will agglomerate as much as required to stay below such limit. Thus, the particle agglomerate size is a direct function of the energy applied per spark and its frequency. The sparking frequency is controlled directly by the current in the circuit. The current does not modify the electric shape of a single spark.

**Gap separation**: the higher the gap, the higher the breakdown voltage required to cause a spark. There seem to be 2 domains: one at gap separations < 3 mm in $N_2$, where the slope of breakdown voltage increases as a function of the gap, and the other for gaps > 3 mm, where the mentioned slope is less pronounced. The maximum in agglomerate size and mass production was obtained at 3 mm. This suggests that the conversion efficiency from spark energy to nanoparticle ablation is a direct function of gap separation, and could be given by the spark diameter.

**Gas flow rate:** starting from around 4 lpm, the higher the flow rate, the smaller the nanoparticles, but also the higher its concentration. Higher flow rates also increase the breakdown voltage.

**Resistance:** at a given resistance, where the stability factor is critically damped, the circuit interferes with the oscillations of the spark and cable, reducing the energy transfer from the circuit into the spark. This produces the smallest and lowest concentration of nanoparticles. In our system, it is observed at ~100 Ω when the lowest number of oscillations in the spark is produced. A larger number of oscillations seems to be linked with larger particles, higher concentration and more energy per spark. Also, the largest resistances increase the frequency as the circuit side does not go down to 0 V after discharge. The simple circuit becomes more complex for the fast spark dynamics. In this work, we have seen that above a certain resistance threshold, it is possible to isolate the main circuit and the only part affecting the spark is the cable after the resistor. In such a way, one can tune the spark towards desirable conditions, either the spark oscillates significantly due to the cable inductance or oscillates less. This information is very useful for systems where nanoparticles are purposefully generated, but it can also be useful for minimizing aerosol formation in applications such as spark plugs used in commercial Otto cycle engines, or any electronic ignition system. In such applications, a dim spark that does not consume the electrode material is more desirable for allowing the longest possible lifetime of the component.

**Capacitance:** it determines the spark energy and its duration. The larger the capacitance, the bigger and more concentrated the nanoparticles are produced. Capacitance is inversely correlated with frequency as the charging of the capacitor takes more time to reach the breakdown voltage. The period of the spark oscillation increases proportionally to the capacitance in the circuit.

**Inductance:** values of inductance required to make the system inductive are typically much smaller than the inductance of the bare cable. All the measurements for induction laid on the underdamped regime.

The resistance values present in commercial inductors and the relatively low saturation currents make it challenging to isolate the pure effect of inductance in the circuit. Some correlations observed are: in general, larger inductances seem to cause more ripples (oscillations). The presence of resistance seems to reduce the amount of these ripples. At certain conditions, the natural oscillation of the spark, cable and circuit interfere with one another and cause irregular but reproducible oscillations (100 µH and 1mH torus). At other spark conditions, the system seems to be critically damped. The observations here also seem to support the idea of having a larger number of oscillations being beneficial for the energy efficiency from the circuit into nanoparticle ablation.

The particle number concentration seems then to be correlated to the effective energy given to the spark, and adjusted by the efficiency of spark energy to nanoparticle evaporation. Some equations have been proposed in the past to estimate the energy available per spark, including capacitance and voltage, but they still cannot predict all of our experimental results, particularly in the case of resistance, gap separation and flow rate. This indicates that there are at least two additional factors that need to be considered: firstly, the efficiency of energy transfer from the circuit into the spark and its ultimate utilization to produce nanoparticles, which might depend on the cross sectional area of the spark. Secondly, the RLC behavior of the circuit, which might act in beneficial or detrimental ways depending on its interaction with the natural oscillations caused by the spark. Perhaps an optimal design lies on an RLC circuit where oscillations can be tuned to achieve high electricity efficiency conversion into ablated nanoparticles. This requires not only accurate knowledge of the values of the RLC circuit, including the spark, cable and electrode material, but also, their time-resolved behavior in MHz domain. Such behavior might differ from the nominal values of the individual components and might not remain constant through the spark. Therefore, an appropriate circuit design and tuning are required to leverage towards scaling up this technique and increasing its production efficiency.

This work provides experimental evidence of non-monotonic behaviors (maxima in variable dependence) of the SDG that had been beyond the experimental regions tested by most of the previous reports. The observations do not necessarily invalidate previously proposed mathematical relationships but suggest that additional correlations need to be taken into consideration for a more accurate representation of the SDG system. More studies are encouraged on these lines to further elucidate their quantitative dependencies and minimize electromagnetic noise produced by SDGs. Finally, this type of generator is of great promise for stable, reliable, pure and well-controlled synthesis of nanoparticles for further processing in the synthesis of nanostructured materials.

# Acknowledgements

We are thankful to José Sánchez del Río Sáez for support in the selection of components and construction of the SDG. The authors are grateful for generous financial support provided by the European Union Horizon 2020 Framework Program under grant agreement 678565 (ERC-STEM), 963912 (ERC-PoC, SINERGY), by MINECO for HYNANOSC (RTI2018-099504-A-C22) and "Comunidad de Madrid" for FotoArt-CM (S2018/NMT-4367). This work was also supported by the Marie Sklodowska Curie Fellowship SUPERYARN under grant number: 101029091

# Supplementary information

Table S1. Nomenclature for describing the system conditions with its abbreviations used through this paper.

| Variable | Abbreviation | Units |
|---|---|---|
| Gap separation | $G$ | (mm) |
| Capacitance | $C$ | (farad) |
| Flow rate | $\dot{V}$ | (lpm) |
| Current | $I$ | (mA) |
| Inductance | $L$ | (henry) |
| Resistor before diode | $R_{bd}$ | ($\Omega$) |
| R after inductor | $R_{ai}$ | ($\Omega$) |
| Electrode material | $E_m$ | (-) |

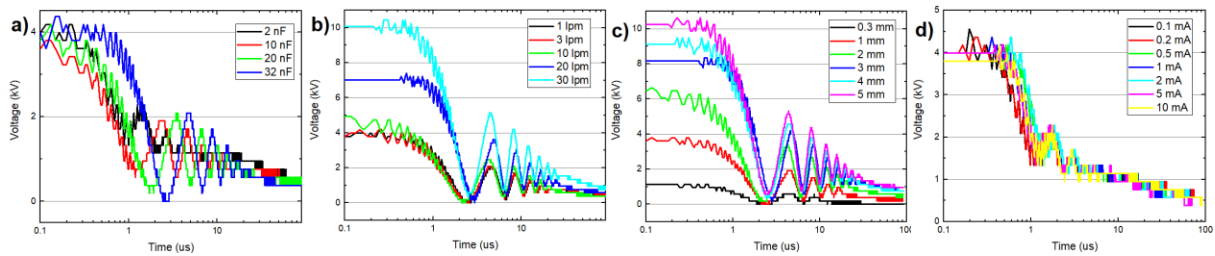

Figure S1. Effect of a) capacitance, b) flow rate, c) gap separation, d) current, on the voltage measured at point $N_1$. Conditions used were: a) $G$=1 mm, $\dot{V}$= 3 lpm, $I$=10 mA. b) $G$=1 mm, $C$=32 nF, $I$=10 mA. c) $C$=32 nF, $\dot{V}$=3 lpm, $I$=10 mA. d) $G$=1 mm, $C$=2 nF, $\dot{V}$=3 lpm. For all cases in this figure: $L$= 1 µH, $R_{bd}$=3 kΩ, $R_{ai}$=none, $E_m$=Fe98%.

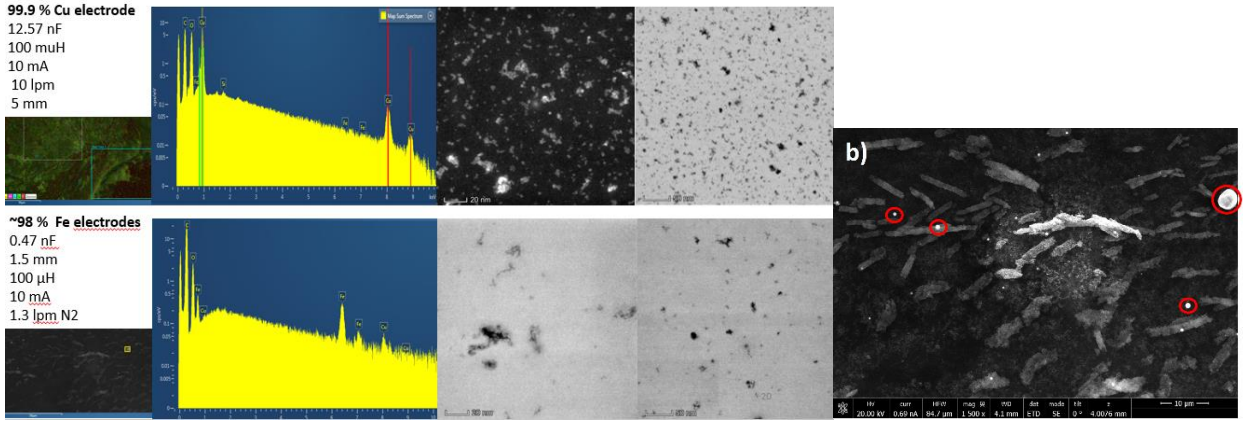

Figure S2. a) Validation of elemental composition of nanoparticles generated using Cu and Fe electrodes via EDX. b) μm sized particles formed due to the recoil ejection of a molten pool of metal in the Fe 98% in the electrode. Some of the micron sized particles have been highlighted for their easier identification in the background of rolled up caked agglomerate of nanoparticles.

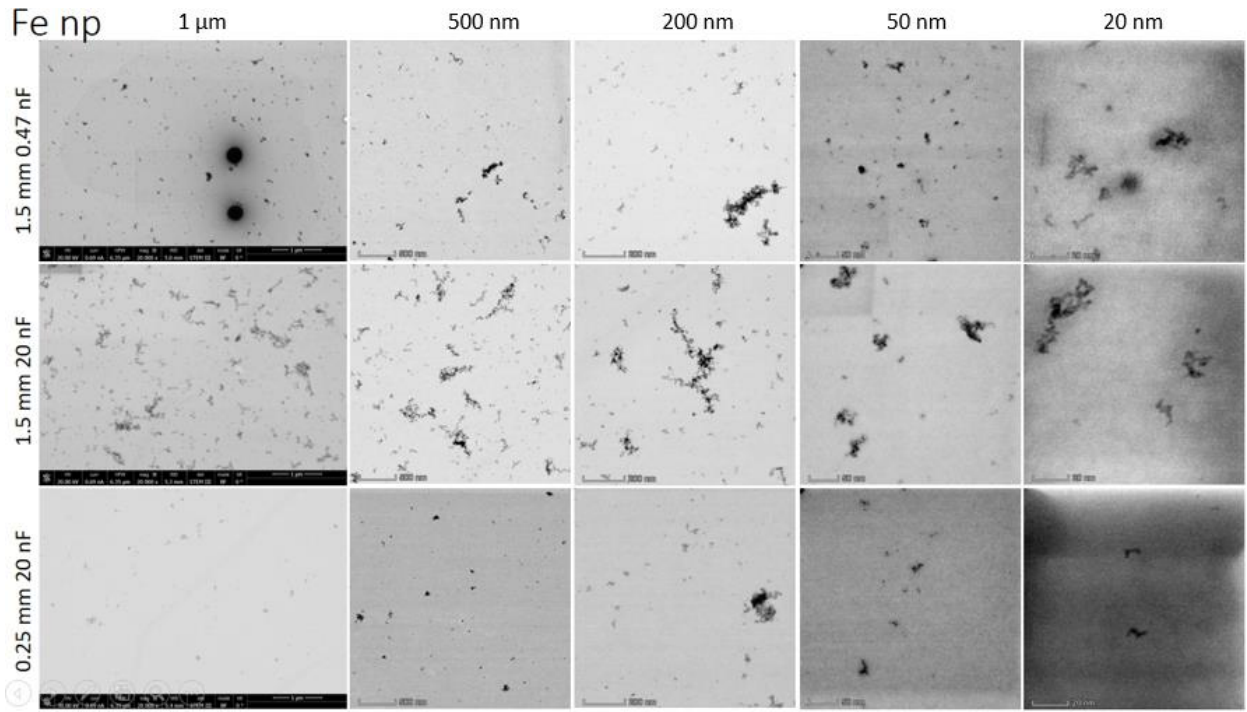

**Figure S3.** Fe nanoparticles showing the effect of different capacitance and gap separation at different magnifications. Measuring band length for each column is on the top. The first column (1 μm scale bar) was taken from the SEM microscope; the other ones were taken from TEM. Conditions used were: G and C as indicated on the left axis, $\dot{V}$=1.3 lpm $H_2$, $I$=5 mA, $L$= 100 μH, $R_{bd}$=3 kΩ, $R_{al}$=none, $E_m$=Fe98%.

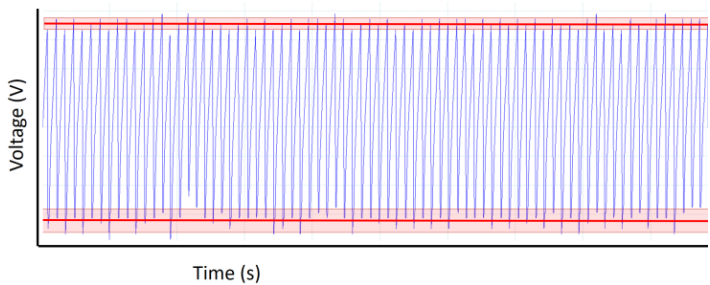

**Figure S4.** Extraction of values of lower mean and upper mean of the spark discharge generator from an oscilloscope. The bounds represent the standard deviation from the upper and the lower peak respectively calculated from a random sample of 10 sparks (full oscillations).

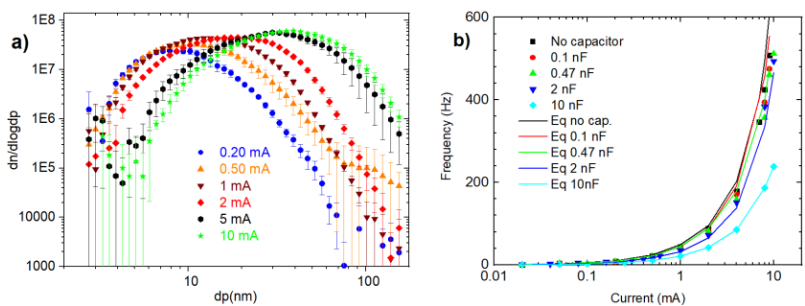

**Figure S5. a)** Size distribution as a function of current for conditions of Figure 3c. **b)** Effect of current on frequency for different capacitances.

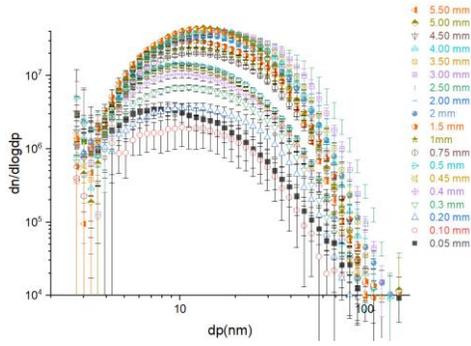

**Figure S6.** Size distributions for the effect of gap distance. Mean and total concentration are given in Figure 4c.

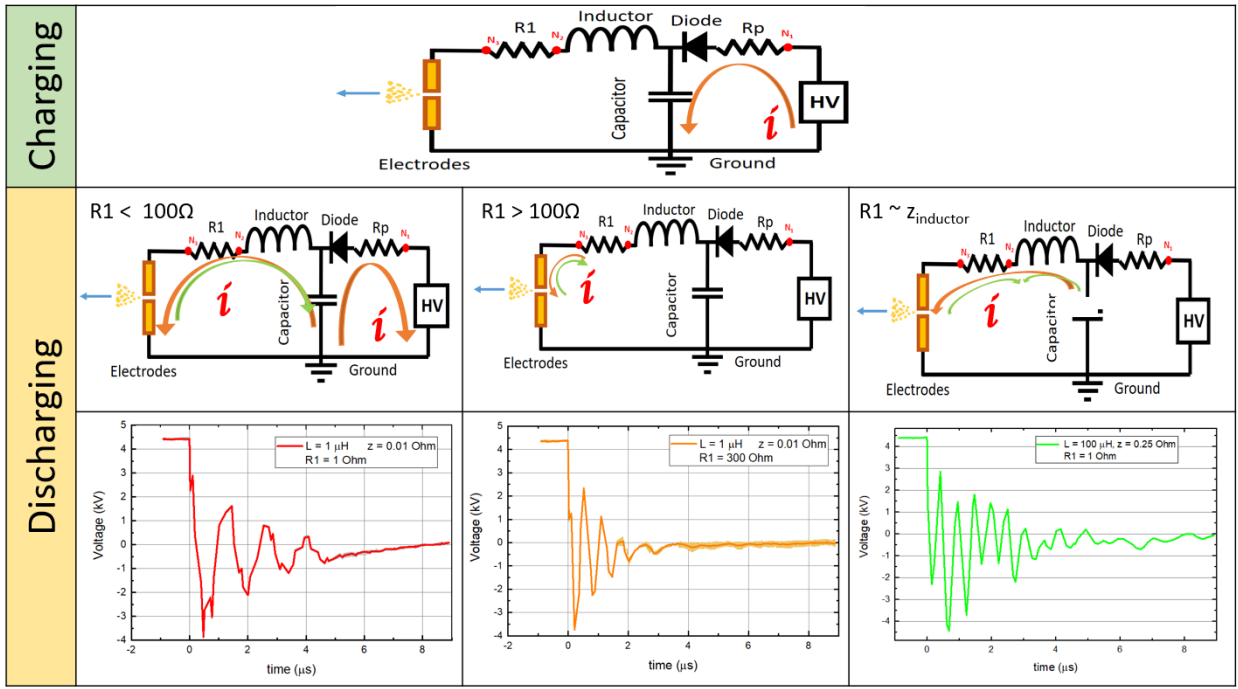

**Figure S7.** RLC circuit topology during charge and discharge as well as part of the circuit that dominates the spark behavior for the first discharge. During oscillation, the current reverses as the voltage changes sign, but it lies within one of the discharging steps shown schematically based on the magnitude of inductance and resistance. In the first condition (R1<100 Ω), the whole circuit plays a role in defining spark features and the oscillation period is the largest. In the second (R1 > 100Ω), only the part after the resistance affects the spark shape and the oscillation period is smaller than in the previous case. In the third case (z=0.25 Ω and R1=1 Ω) the resistance and the inductor seem to interfere with one another. At critical conditions the duration of the oscillation and spark reach a minimum.

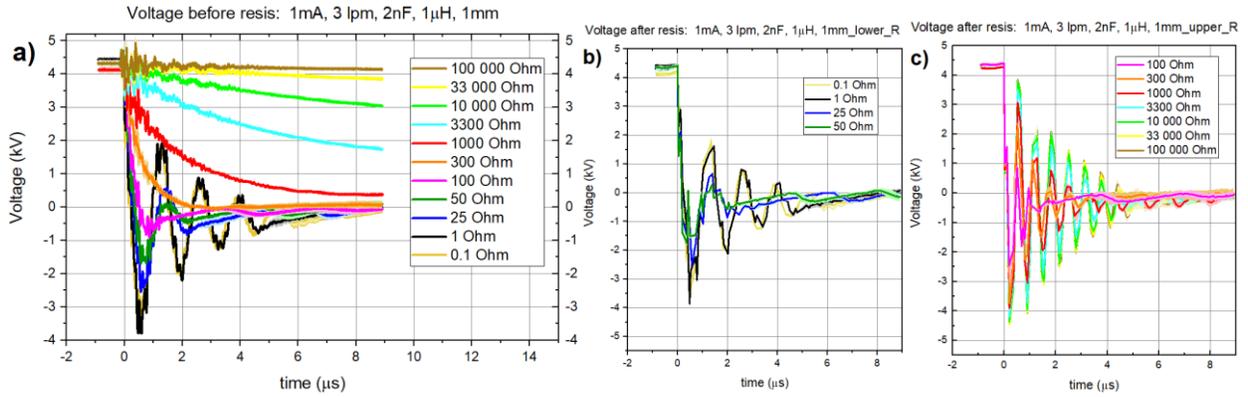

**Figure S8.** 2D representation of the effect of different resistances on the voltage **a)** before and **b)** after them for the resistances <50 Ohms (Underdamped condition) and **c)** after for the resistances ≥ 100 Ohm (Overdamped condition). The same data can be found in the 3D plots of Figure 6.

**Time-resolved energy calculation:**

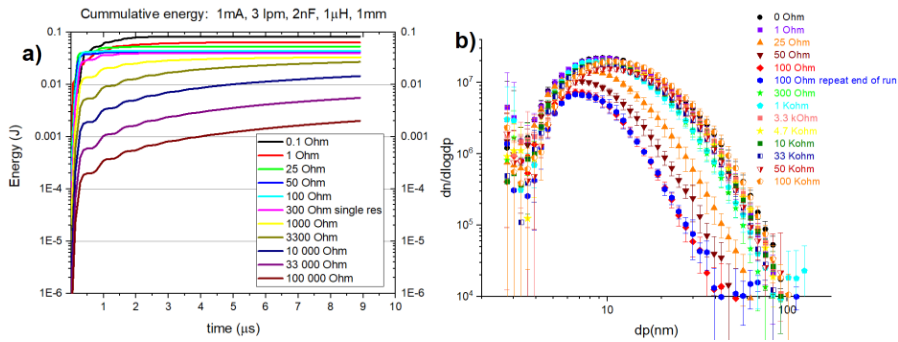

**Figure S9. a)** Time-resolved energy calculation for different values of resistors R1. Since the values for energy have not converged to an asymptote for resistors higher than 3300 Ω at the shown µs regime, they were characterized in ms regime as Figure S10 shows. **b)** Size distribution generated for the different resistances used.

**Resistance effect for ms time resolution:**

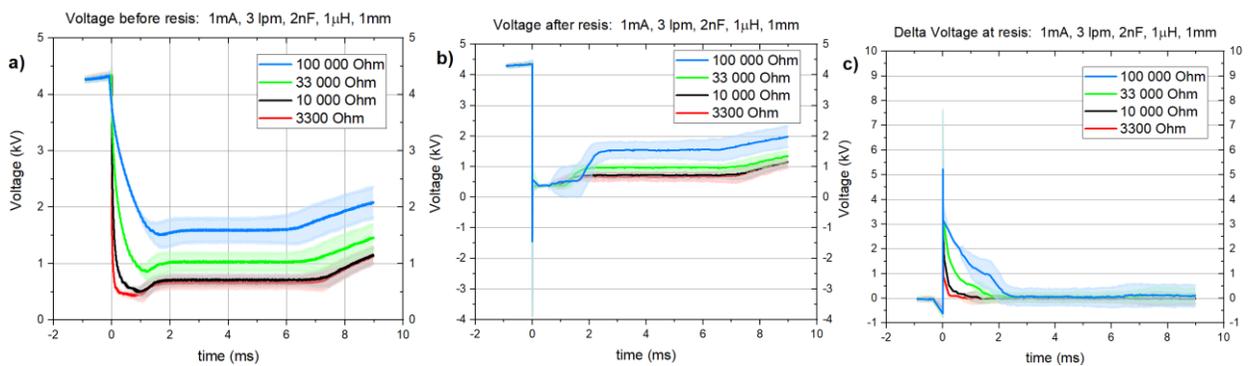

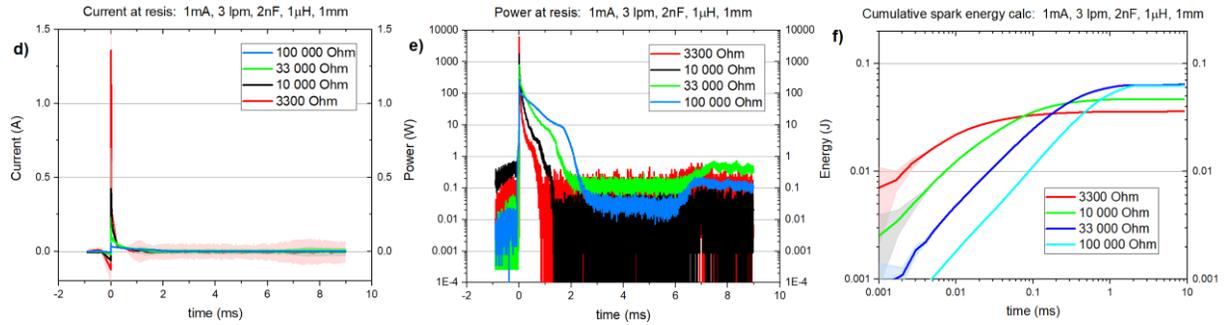

**Figure S10.** Millisecond resolved spark characterizations and calculations for high resistance values: **a)** voltage before R1, **b)** voltage after R1, **c)** delta voltage in resistor **d)** current calculated via Ohm´s law, **e)** power estimated, **f)** cumulative spark energy estimated as calculated from R1 dynamics.

**Measurement error**

Huge error propagation is caused due to the nature of the calculation of current, first kV to V conversion, via the high voltage probe 1000:1 is 3%. The Picoscope has an accuracy of 3% in DC and 4% in AC, and the error of the fixed resistance is nominally 5%. The error tends to be constant and not random in some cases causing the final error after subtraction to be a bit better than this number. Nevertheless, the intrinsic Ohms law for the estimation of current causes the error to become amplified several orders of magnitude for the cases of low resistance. These are challenges of the measurement system and not necessarily represent a disadvantage of the SDG running at such conditions. To deal with such signal noise issues, filtering was used following a criteria: If the standard deviation is higher than 2 times the value of the processed current data, it is taken as a marginally low value that still allows time-resolved logarithmic plotting (such as 0.001 Amp).

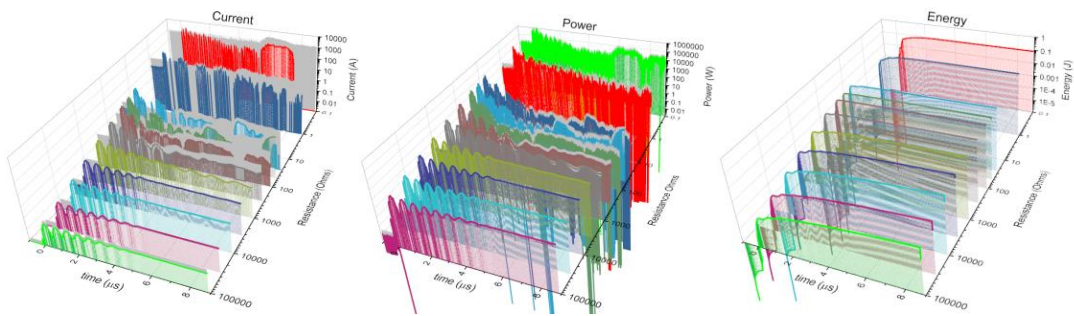

**Figure S11.** Effect of error propagation in current, power and energy calculation as a function of resistance. In this figure, no correction was made for large standard deviations. Compare these figures with those where the criteria for very large standard deviations were used (Figure 6g).

**Variable resistor for studying resistance effect in high voltage applications:**

Care should be taken when using variable resistors as they are not rated for these high voltages and they might short circuit and spark in between the circuitry, causing the spark in the generator to be weaker and display a behavior corresponding to a different condition.

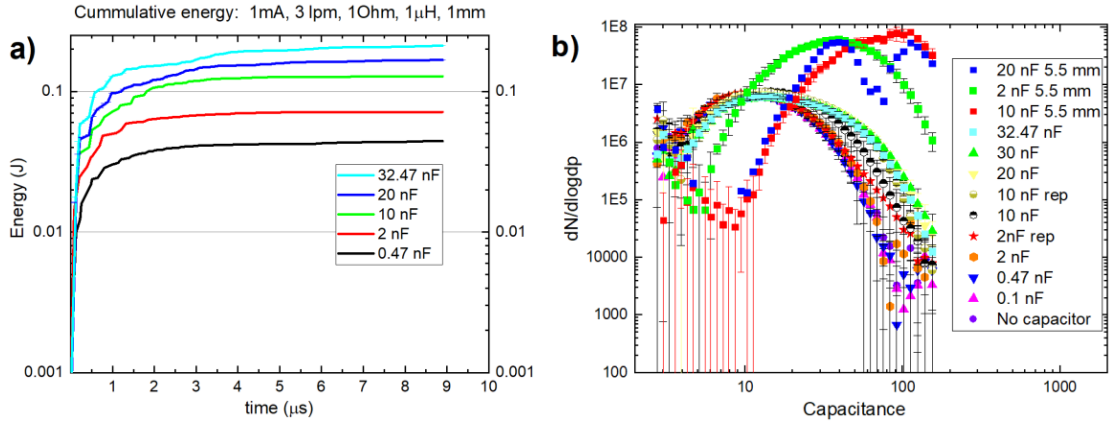

Figure S12. Capacitance effect: a) 2D plot of cumulative energy corresponding to the 3D plot of Figure 7g. b) Size distribution of particles whose geometric mean size and total concentration is plotted in Fig 7b. (Those without a specified gap it is 1 mm).

**Inductance effect**

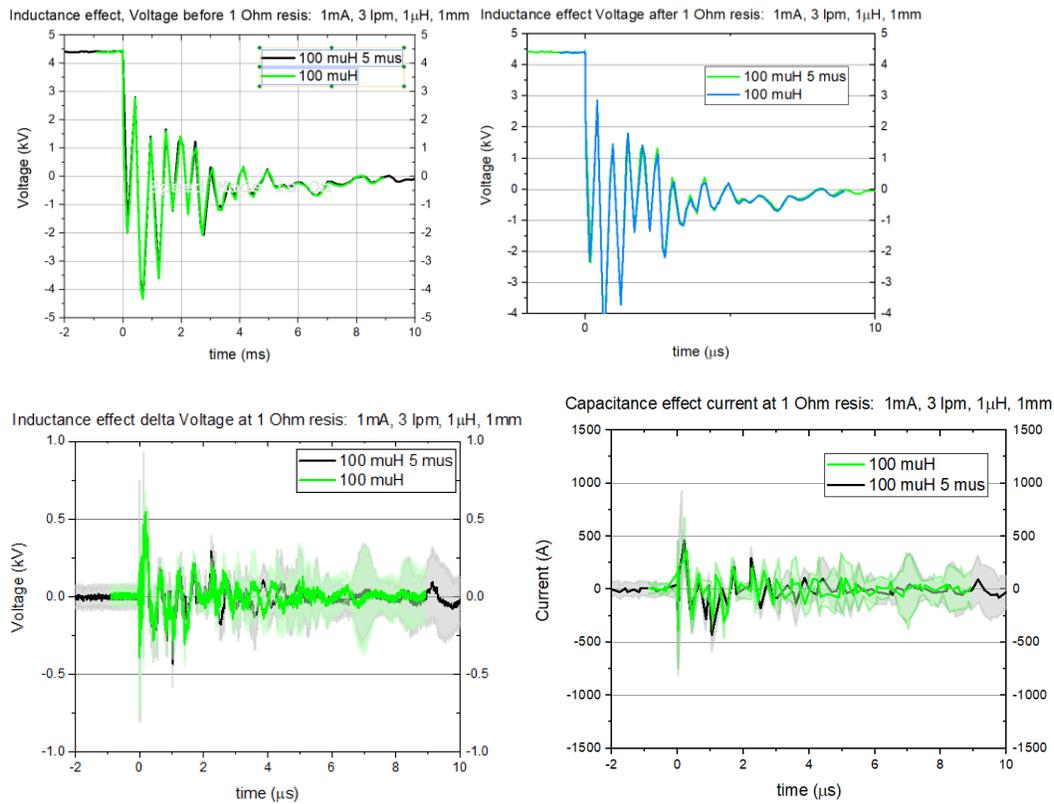

Figure S13. Effect on length of data acquired for the 100 µH inductor. Signal is quite reproducible.

**Inductance effect additional data:**

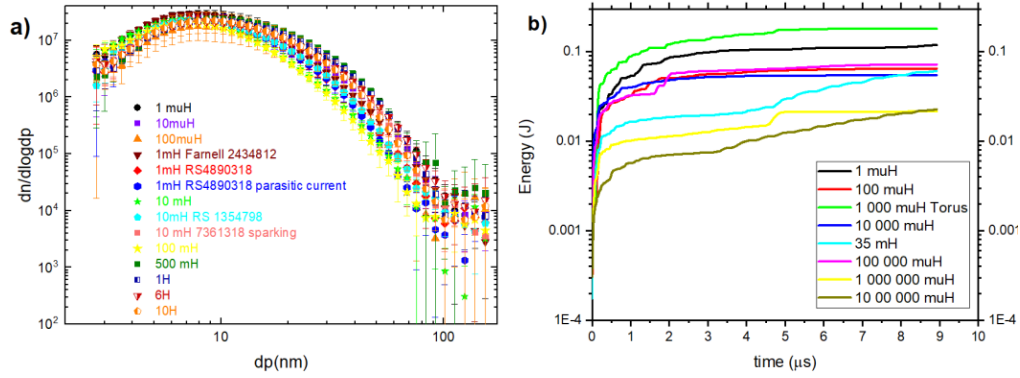

Figure S14. a) Particle size distribution for different inductors. b) Inductance effect on energy per spark calculation.

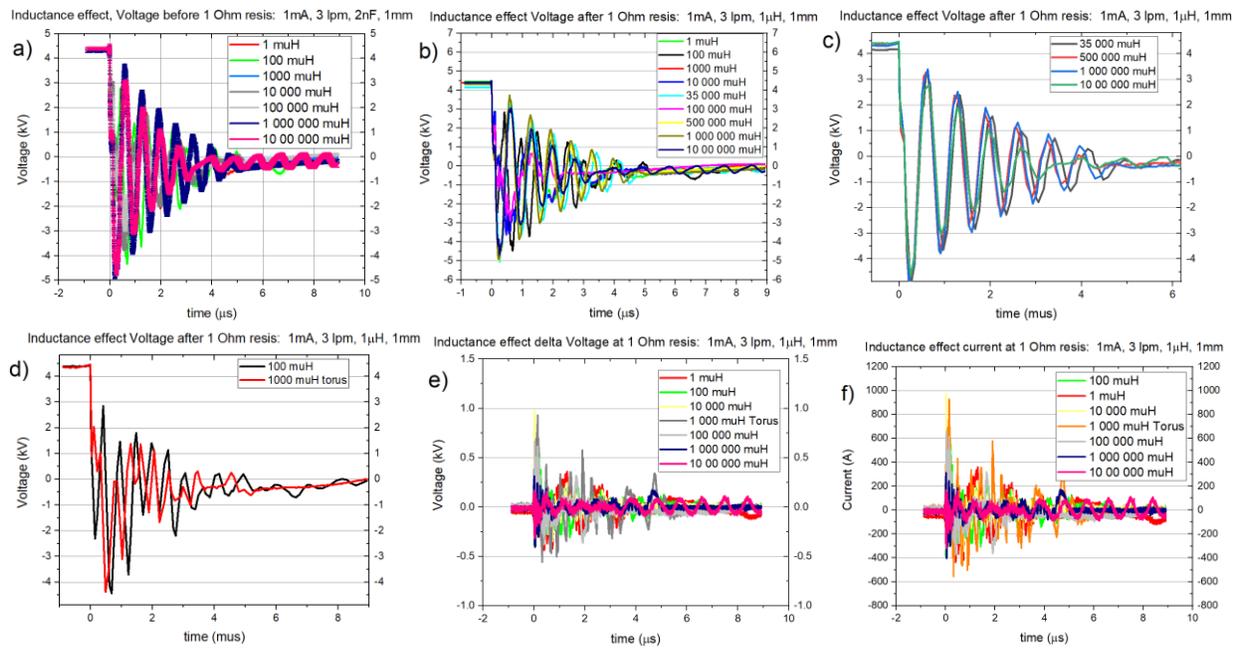

Figure S15. a) 2D version of voltage before and b) after resistor R1 for inductance effect.

**Millisecond resolved data:**

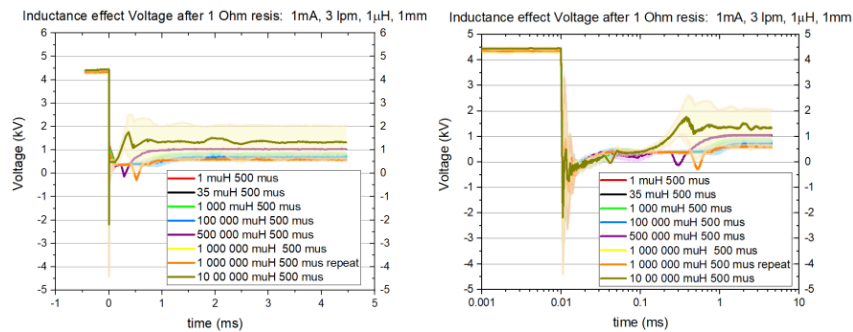

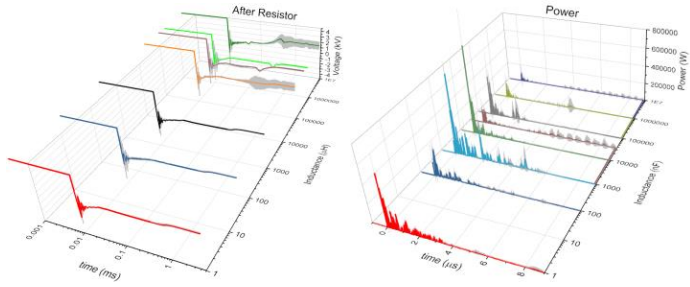

**Figure S16. a)** Shoulder observation on the effect of inductance at ms regime. Time-resolved power calculated for inductance effect. **b)** Transitory power estimated via Eq. 16.

Table S2. Specs of Inductors utilized.

| Inductance | Product No | RS /Farnell | Resistance max (Ohm) | Sat. current (mA) |
|---|---|---|---|---|
| 1 µH | 1635794 | Farnell | 0.01 Ω | 10 A |
| 10 µH | 2965268 | Farnell | 450 mΩ | 5 A |
| 100 µH | 13R104C 2062696 | Farnell With 104 C on top | 0.24 Ω | 1 A |
| 1 mH | 7361882 | RS | 6 Ω | 270 mA |
| 1 mH | 7409520 | RS | 1.8 Ω | 1 A |
| 1 mH | 489-0318 | RS Wurth, 1 mH (torus) | 7 mΩ | 10 A |
| 1 mH | 2434812 | Farnell | 3.9 Ω | 120 mA |
| 10 mH | 7361318 | RS | 39 Ω | 100 mA |
| 10 mH | 1354798 | RS | 22.8 Ω | 130 mA |
| 35 mH | 1834333   C-66U | Farnell | 0.79 Ω | 2 A |
| 100 mH | 2725293 | Farnell  RLB0913-104K | 235 Ω | 20 mA |
| 500 mH | C-36X   1838355 | Farnell | 30 Ω | 300 mA |
| 1 H | C-24X   1834830 | Farnell | 50 Ω | 240 mA |
| 6 H | C-14X   1834331 | Farnell | 150 Ω | 200 mA |
| 10 H | C-3X | Farnell | 500 Ω | 50 mA |

Table S3. Inductor behavior categorization as explained by an effective inductance term given by multiplying nominal inductance times saturation ratio and divided by a hypothetical spark with a current of 200 A. The behavior as a whole circuit interaction is given for effective inductances ($H_{eff}$) of 0.05–5 µH, mixed signals by $H_{eff}$ of 0.5–50 µH, critical conditions with $H_{eff}$=10 µH and largest inductor case for $H_{eff}$ larger than 250 µH.

| Inductance (µH) | Resistance (Ω) | Oscillations per spark (No.) | Saturation current (A) | Effective inductance (µH) | Description |
|---|---|---|---|---|---|
| 1 | 0.01 | 3.5 | 10 | 0.05 | Whole circuit interaction |
| 100 | 0.24 | 9 | 1 | 0.5 | mixed signal |
| 1000 | 1.4 | 2.5 | 0.9 | 4.5 | Whole circuit interaction |
| 1000 | 0.007 | 8 | 10 | 50 | Mixed signal |
| 10000 | 39 | 2.5 | 0.1 | 5 | Whole circuit interaction |
| 35000 | 0.79 | 7 | 2 | 350 | Largest inductors case |
| 100000 | 235 | 2 | 0.02 | 10 | Critical condition |
| 500000 | 30 | 6 | 0.3 | 750 | Largest inductors case |
| 1000000 | 50 | 6.5 | 0.05 | 250 | Largest inductors case |
| 10000000 | 500 | 4 | 0.05 | 2500 | Largest inductors case |